\documentclass{article}
\usepackage{graphicx}
\usepackage{hyperref}
\usepackage{amsmath}
\usepackage{amsfonts}
\usepackage{amssymb}

\providecommand{\U}[1]{\protect\rule{.1in}{.1in}}

\input{tcilatex}

\begin{document}

\author{Antony Valentini \\
Augustus College}

\begin{center}
{\large Foundations of statistical mechanics and the status of the Born rule
in de Broglie-Bohm pilot-wave theory}

\bigskip

\bigskip

\bigskip

Antony Valentini

\textit{Augustus College,}

\textit{14 Augustus Road,}

\textit{London SW19 6LN, UK.}

\textit{Department of Physics and Astronomy,}

\textit{Clemson University, Kinard Laboratory,}

\textit{Clemson, SC 29634-0978, USA.}

\bigskip

\bigskip
\end{center}

\bigskip

\bigskip

We compare and contrast two distinct approaches to understanding the Born
rule in de Broglie-Bohm pilot-wave theory, one based on dynamical relaxation
over time (advocated by this author and collaborators) and the other based
on typicality of initial conditions (advocated by the `Bohmian mechanics'
school). It is argued that the latter approach is inherently circular and
physically misguided. The typicality approach has engendered a deep-seated
confusion between contingent and law-like features, leading to misleading
claims not only about the Born rule but also about the nature of the wave
function. By artificially restricting the theory to equilibrium, the
typicality approach has led to further misunderstandings concerning the
status of the uncertainty principle, the role of quantum measurement theory,
and the kinematics of the theory (including the status of Galilean and
Lorentz invariance). The restriction to equilibrium has also made an
erroneously-constructed stochastic model of particle creation appear more
plausible than it actually is. To avoid needless controversy, we advocate a
modest `empirical approach' to the foundations of statistical mechanics. We
argue that the existence or otherwise of quantum nonequilibrium in our world
is an empirical question to be settled by experiment.

\begin{center}
\bigskip
\end{center}

\bigskip

\bigskip

\bigskip

\bigskip

\bigskip

To appear in: \textit{Statistical Mechanics and Scientific Explanation:
Determinism, Indeterminism and Laws of Nature}, ed. V. Allori (World
Scientific).

\bigskip

\bigskip

\bigskip

\bigskip

\bigskip

\bigskip

\bigskip

\bigskip

\bigskip

\bigskip

\bigskip

\bigskip

\bigskip

\bigskip

\bigskip

\bigskip

\bigskip

\section{Introduction}

The pilot-wave theory of de Broglie (1928) and Bohm (1952a,b) is a
deterministic theory of motion for individual systems. In the version first
given by de Broglie, a system with configuration $q$ and wave function $%
\psi(q,t)$ has an actual trajectory $q(t)$ determined by de Broglie's
equation of motion%
\begin{equation}
\frac{dq}{dt}=\frac{j}{\left\vert \psi\right\vert ^{2}}\ ,   \label{deB}
\end{equation}
where $\psi$ obeys the usual Schr\"{o}dinger equation (units $\hbar=1$)%
\begin{equation}
i\frac{\partial\psi}{\partial t}=\hat{H}\psi   \label{Sch}
\end{equation}
(with a Hamiltonian operator $\hat{H}$) and $j$ is a current satisfying the
continuity equation%
\begin{equation}
\frac{\partial\left\vert \psi\right\vert ^{2}}{\partial t}+\partial_{q}\cdot
j=0   \label{Cont1}
\end{equation}
(with $\partial_{q}$ the gradient operator in configuration space).\footnote{%
This construction applies to any system with a Hamiltonian $\hat{H}$ given
by a differential operator on configuration space (Struyve and Valentini
2009).} Equation (\ref{Cont1}) is a straightforward consequence of (\ref{Sch}%
), and using (\ref{deB}) it may be rewritten as%
\begin{equation}
\frac{\partial\left\vert \psi\right\vert ^{2}}{\partial t}%
+\partial_{q}\cdot(\left\vert \psi\right\vert ^{2}\dot{q})=0   \label{Cont2}
\end{equation}
(where $\dot{q}=j/\left\vert \psi\right\vert ^{2}$ is the
configuration-space velocity field).

Thus, for example, for a single low-energy spinless particle of mass $m$ we
find a current%
\begin{equation}
\mathbf{j}=\left\vert \psi\right\vert ^{2}\frac{\mathbf{\nabla}S}{m}
\end{equation}
(where $S$ is the phase of $\psi=\left\vert \psi\right\vert \exp(iS)$) and (%
\ref{deB}) reads%
\begin{equation}
\frac{d\mathbf{x}}{dt}=\frac{\mathbf{\nabla}S}{m}\ .   \label{deB_1part}
\end{equation}

Given an initial wave function $\psi(q,0)$, (\ref{Sch}) determines $\psi(q,t)
$ at all times and so the right-hand side of (\ref{deB}) is also determined
at all times. Given an initial configuration $q(0)$, (\ref{deB}) then
determines the trajectory $q(t)$. Thus, for example, in a two-slit
experiment with a single particle, if the incident wave function is known
then (\ref{deB_1part}) determines the trajectory $\mathbf{x}(t)$ for any
initial position $\mathbf{x}(0)$.

Mathematically, for a given wave function, the law of motion (\ref{deB})
defines a trajectory $q(t)$ for each initial configuration $q(0)$. In
practice, however, we do not know the value of $q(0)$ within the initial
packet. For an ensemble of systems with the same $\psi(q,0)$, the value of $%
q(0)$ will generally vary from one system to another. We may then consider
an initial distribution $\rho(q,0)$ of values of $q(0)$ over the ensemble.
As the trajectories $q(t)$ evolve, so will the distribution $\rho(q,t)$. By
construction $\rho(q,t)$ will obey the continuity equation%
\begin{equation}
\frac{\partial\rho}{\partial t}+\partial_{q}\cdot(\rho\dot{q})=0\ . 
\label{Cont3}
\end{equation}

In principle there is no reason why we could not consider an arbitrary
initial distribution $\rho(q,0)$. De Broglie's equation (\ref{deB})
determines the time evolution of a trajectory $q(t)$ for any initial $q(0)$,
and over the ensemble the continuity equation (\ref{Cont3}) determines the
time evolution of a density $\rho(q,t)$ for any initial $\rho(q,0)$. There
is certainly no reason of principle why $\rho(q,0)$ should be equal to $%
\left\vert \psi(q,0)\right\vert ^{2}$.

As an extreme example, an ensemble of one-particle systems could have the
initial distribution $\rho(\mathbf{x},0)=\delta^{3}(\mathbf{x}-\mathbf{x}%
_{0})$, with every particle beginning at the same point $\mathbf{x}_{0}$. As
the distribution evolves, it will remain a delta-function concentrated on
the evolved point $\mathbf{x}(t)$. Every particle in the ensemble would
follow the same trajectory. If such an ensemble were fired at a screen with
two slits, every particle would land at the same final point $\mathbf{x}_{f}$
on the backstop and there would be no interference pattern (indeed no
pattern at all), in gross violation of quantum mechanics.

If instead it so happens that $\rho(q,0)=\left\vert \psi(q,0)\right\vert ^{2}
$, then since $\rho$ and $\left\vert \psi\right\vert ^{2}$ obey identical
continuity equations ((\ref{Cont3}) and (\ref{Cont2}) respectively) it
follows that%
\begin{equation}
\rho(q,t)=\left\vert \psi(q,t)\right\vert ^{2}   \label{Born}
\end{equation}
for all $t$. This is the usual Born rule. In conventional quantum mechanics (%
\ref{Born}) is taken to be an axiom or law of nature. Whereas in pilot-wave
theory (\ref{Born}) is a special state of `quantum equilibrium': if it
happens to hold at one time it will hold at all times (for an ensemble of
isolated systems). Thus, for example, if such an ensemble of particles is
fired at a screen with two slits, the incoming equilibrium ensemble will
evolve into an equilibrium ensemble at the backstop, and hence the usual
interference pattern $\rho=\left\vert \psi\right\vert ^{2}$ will be
trivially obtained. More generally, as first shown in detail by Bohm
(1952b), for systems and apparatus initially in quantum equilibrium, the
distribution of outcomes of quantum measurements will agree with the
conventional quantum formalism.

But in principle the theory allows for `quantum nonequilibrium' ($\rho
\neq\left\vert \psi\right\vert ^{2}$). How then can pilot-wave theory
explain the success of the Born rule (\ref{Born}), which has been confirmed
to high accuracy in every laboratory experiment? Most workers in the field
(past and present) simply take it as a postulate. Thus, for example,
according to Bell (1987, p. 112) `[i]t is \textit{assumed} that the
particles are so delivered initially by the source', while according to
Holland (1993, p. 67) the Born rule is one of the `basic postulates'. This
is unsatisfactory. There is after all a basic conceptual distinction between
equations of motion and initial conditions. The former are regarded as
immutable laws (they could not be otherwise), whereas the latter are
contingencies (there is no reason of principle why they could not be
otherwise). Once the laws are known they are the same for all systems,
whereas for a given system the initial conditions must be determined
empirically. Thus Newton, for example, wrote down laws that explain the
motion of the moon, but he made no attempt to explain the current position
and velocity of the moon: the latter are arbitrary or contingent initial
conditions to be determined empirically, which may then be inserted into the
laws of motion to determine the position and velocity at other times. In
pilot-wave theory, if we consider only ensembles restricted by the
additional postulate (\ref{Born}), then this is closely analogous to
considering Newtonian mechanics only for ensembles restricted to thermal
equilibrium (with a uniform distribution on the energy surface in phase
space). In both theories there is a much wider nonequilibrium physics, which
is lost if we simply adopt initial equilibrium as a postulate.

Most workers in the field seem unperturbed by this and continue to treat (%
\ref{Born}) as a postulate. Others are convinced that some further
explanation is required and that the question -- if pilot-wave theory is
true, why do we always observe the Born rule? -- requires a more satisfying
answer.

There are currently two main approaches to understanding the Born rule in
pilot-wave theory, which we briefly summarise here.

The first approach, associated primarily with this author and collaborators,
proposes that the Born rule we observe today should be explained by a
process of `quantum relaxation' (analogous to thermal relaxation), whereby
initial nonequilibrium distributions $\rho\neq\left\vert \psi\right\vert ^{2}
$ evolve towards equilibrium on a coarse-grained level, $\bar{\rho}%
\rightarrow \overline{\left\vert \psi\right\vert ^{2}}$ (in terms of
coarse-grained densities $\bar{\rho}$ and $\overline{\left\vert
\psi\right\vert ^{2}}$, much as in Gibbs' classical account of thermal
relaxation for the coarse-grained density on phase space). This process may
be understood in terms of a `subquantum' coarse-graining $H$-theorem on
configuration space, analogous to the classical coarse-graining $H$-theorem
on phase space (Valentini 1991a,b). Extensive numerical simulations, carried
out with wave functions that are superpositions of different energy states,
have confirmed the general expectation that initial densities $\rho(q,0)$
lacking in fine-grained microstructure rapidly become highly filamentary on
small scales and indeed approach the equilibrium density $\left\vert
\psi\right\vert ^{2}$ on a coarse-grained level (Valentini 1992, 2001;
Valentini and Westman 2005; Towler, Russell and Valentini 2012; Colin 2012;
Abraham, Colin and Valentini 2014). This may be quantified by a decrease of
the coarse-grained $H$-function%
\begin{equation}
\bar{H}(t)=\int dq\ \bar{\rho}\ln\left( \bar{\rho}/\overline{\left\vert
\psi\right\vert ^{2}}\right) \ ,   \label{Hbar}
\end{equation}
which reaches its minimum $\bar{H}=0$ if and only if $\bar{\rho}=\overline{%
\left\vert \psi\right\vert ^{2}}$, and which is found to decay approximately
exponentially with time (Valentini and Westman 2005; Towler, Russell and
Valentini 2012; Abraham, Colin and Valentini 2014). Similar studies and
simulations have been carried out for field theory in an expanding universe,
for which relaxation is found to be suppressed at very long cosmological
wavelengths (Valentini 2007, 2008a, 2010a; Colin and Valentini 2013, 2015,
2016). This opens the door to possible empirical evidence for quantum
nonequilibrium in the cosmic microwave background (Valentini 2010a; Colin
and Valentini 2015; Vitenti, Peter and Valentini 2019) -- as well as in
relic particles left over today from the very early universe (Valentini
2001, 2007; Underwood and Valentini 2015, 2016). It has further been shown
that if nonequilibrium systems were discovered today, their physics would be
radically different from the physics we currently know, involving practical
superluminal signalling, violations of the uncertainty principle, and a
general breakdown of standard quantum constraints (such as expectation
additivity and the indistinguishability of non-orthogonal quantum states)
(Valentini 1991a,b, 1992, 2002a, 2004, 2009; Pearle and Valentini 2006). On
this view, quantum theory is merely a special `equilibrium' case of a much
wider nonequilibrium physics, which may have existed in the early universe
and which could still exist in some exotic systems today.

The second approach, associated primarily with D\"{u}rr, Goldstein, and Zangh%
\`{\i}, as well as with Tumulka and other collaborators, proposes that the
Born rule we observe today should be explained in terms of the `typicality'
of configurations $q_{\mathrm{univ}}(0)$ for the whole universe at the
initial time $t=0$ (D\"{u}rr, Goldstein and Zangh\`{\i} 1992; D\"{u}rr and
Teufel 2009; Goldstein 2017; Tumulka 2018). In this approach, if $\Psi_{%
\mathrm{univ}}(q_{\mathrm{univ}},0)$ is the initial universal wave function
then $|\Psi_{\mathrm{univ}}(q_{\mathrm{univ}},0)|^{2}$ is assumed to be the
natural measure on the set of possible initial universal configurations $q_{%
\mathrm{univ}}(0)$. It may then be shown that the Born rule (\ref{Born}) is
almost always obtained for ensembles of sub-systems prepared with wave
function $\psi$ -- where `almost always' is defined with respect to the
measure $|\Psi_{\mathrm{univ}}(q_{\mathrm{univ}},0)|^{2}$. This is regarded
as an explanation for the empirical success of the Born rule (\ref{Born}).
On this view there is no realistic chance of ever observing quantum
nonequilibrium, which is intrinsically unlikely (as defined with respect to $%
|\Psi_{\mathrm{univ}}(q_{\mathrm{univ}},0)|^{2}$). The Born rule is in
effect regarded as an intrinsic part of the theory, though instead of
postulating the probability distribution (\ref{Born}) for sub-systems this
approach postulates the typicality measure $|\Psi_{\mathrm{univ}}(q_{\mathrm{%
univ}},0)|^{2}$ for the whole universe at $t=0$. If this is correct, quantum
nonequilibrium will never be observed and de Broglie-Bohm theory will never
be experimentally distinguishable from conventional quantum theory.

The typicality approach has given rise to a distinctive physical perspective
on pilot-wave theory -- concerning for example the status of the uncertainty
principle and of Lorentz invariance, among other important topics. These
views may be classified under the heading of the `Bohmian mechanics school',
where the term `Bohmian mechanics' was first introduced by D\"{u}rr,
Goldstein and Zangh\`{\i} (1992) to denote the dynamical theory defined by
equations (\ref{deB}) and (\ref{Sch}).

It should however be noted that, historically speaking, the dynamics defined
by (\ref{deB}) and (\ref{Sch}) was first proposed by de Broglie at the 1927
Solvay conference (for a many-body system with a pilot wave in configuration
space) (Bacciagaluppi and Valentini 2009). De Broglie called his new form of
dynamics `pilot-wave theory'. The theory was revived by Bohm in 1952, though
rewritten in a second-order form with a law of motion for acceleration that
includes a `quantum potential' $Q$. Bohm's version of the dynamics is
physically distinct from de Broglie's:\textbf{\ }in principle it allows for
non-standard initial momenta $p\neq\partial_{q}S$ (Bohm 1952a, pp. 170, 179;
Colin and Valentini 2014).\footnote{%
In Bohm's dynamics there arises the additional question of why we observe
today an `extended quantum equilibrium' in phase space, with momenta
satisfying $p=\partial_{q}S$ as well as configurations distributed according
to (\ref{Born}). Colin and Valentini (2014) show that extended
nonequilibrium does not relax and is unstable, and argue that as a result
Bohm's dynamics is physically untenable.} Thus there are important physical
differences between de Broglie's dynamics and Bohm's dynamics. The
terminology `Bohmian mechanics', as applied to de Broglie's equations (\ref%
{deB}) and (\ref{Sch}), is therefore misleading: it does not give due credit
to de Broglie and it misrepresents the views of Bohm. Our concern in this
paper is with the status of the Born rule in de Broglie's dynamics which,
following de Broglie's own usage (as well as Bell's), we refer to as
`pilot-wave theory'.

Writings by the Bohmian mechanics school generally fail to recognise the
significance and priority of de Broglie's work. For example, in rather glib
sections entitled `History', both Goldstein (2017)\ and Tumulka (2018)
portray de Broglie as having simply proposed or considered the guidance
equation at the 1927 Solvay conference. But in fact, as early as 1923 de
Broglie had postulated the single-particle guidance equation -- as a new law
of motion, expressing a unification of the principles of Maupertuis and
Fermat -- and in that same year de Broglie used his theory to predict
electron interference (four years before it was observed by Davisson and
Germer). Furthermore, it was de Broglie's early research into his new form
of dynamics (with particles guided by waves) that led Schr\"{o}dinger to the
wave equation in 1926.\footnote{%
For an extensive historical analysis of de Broglie's remarkable work in the
period 1923--27, see Bacciagaluppi and Valentini (2009, chapter 2).}

The distinctive approach of the Bohmian mechanics school has been reiterated
and developed in a number of papers and reviews. A textbook has also been
published (D\"{u}rr and Teufel 2009). For the sake of perspective it is
worth remarking that writings by members of this school generally focus on
their own interpretation. There are other approaches to de Broglie-Bohm
theory, not only that taken by this author and collaborators but also others
that lie outside the scope of this paper.\footnote{%
See, for example, the books by Holland (1993)\ and by Bohm and Hiley (1993).}
The Bohmian mechanics school has been particularly influential among
philosophers of physics. The entry `Bohmian mechanics' in \textit{The
Stanford Encyclopedia of Philosophy} is written by a leading member of the
school (Goldstein 2017) (regularly updated by the same author since 2001).
It is noteworthy that such an extensive reference encyclopedia does not
contain an entry on de Broglie-Bohm theory generally; only this one
particular school is represented, suggesting a skewed perception of the
field among philosophers. One of the aims of this paper is to redress this
imbalance in the philosophy of physics literature.

We shall compare and contrast the two approaches outlined above, in
particular regarding the status of the Born rule and related physical
questions. As we shall discuss, in our view the typicality approach is
essentially circular (Valentini 1996, 2001). With respect to a different
initial measure (such as $\left\vert \Psi_{\mathrm{univ}}(q_{\mathrm{univ}%
},0)\right\vert ^{4}$), we will almost always obtain initial violations of
the Born rule (such as $\rho\propto\left\vert \psi\right\vert ^{4}$). While
it may be said that nonequilibrium is `untypical' (has zero measure) with
respect to the univeral Born-rule measure, it may equally be said that
nonequilibrium is `typical' (has unit measure) with respect to a
non-Born-rule measure. In effect, in the typicality approach the Born rule
is taken as an axiom, albeit at the level of the universe as a whole. This
is misleading, not least because a postulate about initial conditions can
have no fundamental status in a theory of dynamics.

As we shall see, the typicality approach has engendered a basic confusion
between contingent and law-like features. This has led to misleading claims
not only about the Born rule but also about the nature of the wave function
(or pilot wave). The artificial restriction to equilibrium has led to
further misunderstandings concerning the status of the uncertainty
principle, the role of quantum measurement theory, and the kinematics of the
theory (including the status of Galilean and Lorentz invariance). The
restriction to equilibrium has also made an erroneously-constructed
stochastic model of particle creation seem more plausible than it actually
is.

By considering how hidden variables can account for the Born rule (\ref{Born}%
), workers in quantum foundations find themselves confronted by issues in
the foundations of statistical mechanics -- a subject which is no less
fractious and controversial than quantum foundations itself. We begin by
outlining our own views on the subject (Section 2), summarise some of the
key results in quantum relaxation and how these apply to cosmology (Sections
3 and 4), and then provide a critique of the typicality approach (Section 5)
and of related viewpoints (Sections 6 and 7), ending with some concluding
remarks (Section 8).

\section{Empirical approach to statistical mechanics}

Pilot-wave dynamics is a deterministic theory of motion. As in classical
physics, there is a clear conceptual distinction between the laws of motion
on the one hand and initial conditions\ on the other. The initial conditions
(for the wave function $\psi$ and for the configuration $q$) are in
principle arbitrary -- which is to say, perhaps more properly, that they are
contingent. Whatever the actual initial conditions were, there is no known
reason of principle why they could not have been different. In order to find
out what the initial conditions actually were, we do not appeal to laws or
principles but to simple empiricism: we carry out observations today and on
that basis we try (using our knowledge of the dynamical laws) to deduce the
initial conditions. This is as true for ensembles as it is for single
systems. In pilot-wave theory, for an ensemble of systems with the same
initial wave function $\psi(q,0)$, the initial distribution $\rho(q,0)$ of
actual initial configurations $q(0)$ could in principle be anything. To find
out what $\rho(q,0)$ was in an actual case we must resort to empirical
observation.

This raises the subtle question of what it might mean for pilot-wave
dynamics to `explain' the Born rule (\ref{Born}). If the distribution $\rho$
is purely empirical, there might then seem to be no question of `explaining'
the particular distribution (\ref{Born}): it is not something one explains,
it is something one finds empirically.

The matter is further complicated by the time-reversal invariance of
pilot-wave dynamics. If all initial conditions are in principle possible,
then any distribution $\rho(q,t_{0})$ today is in principle possible, since
it could have evolved from some appropriate $\rho(q,0)$ (which can in
principle be calculated from $\rho(q,t_{0})$ by time-reversal of the
equations of motion). Again, there might seem to be no question of
`explaining' (\ref{Born}): it is simply a brute `matter of fact' established
by observation.

In our view there is in fact considerable scope for explaining the
presently-observed Born rule (\ref{Born}), in the sense that it can be
explained in terms of past conditions (with the aid of dynamical laws) --
where the past conditions are, however, ultimately empirical and not fixed
by any fundamental laws or principles. On this view, the present is
explained in terms of the past, while the past is itself something we
establish empirically. This of course leaves open the possibility of further
explanation by peering even further into the past, and in our universe this
chain of causal explanations eventually leads us back to the big bang. To
understand the origins of the Born rule, then, we are led to consider
conditions in the very early universe. Specifically: what initial conditions
(at or close to the big bang) could have given rise to the all-pervasive
distribution (\ref{Born}) which we see today?

In the context of statistical mechanics, it might be objected that to
explain the state (\ref{Born}) seen today we must not merely deduce which
past conditions (for example, which particular initial states $\rho
(q,0)\neq \left\vert \psi (q,0)\right\vert ^{2}$) could have evolved into (%
\ref{Born}) today, we must instead show that `all' or `most' past conditions
evolve into (\ref{Born}) today. For otherwise, it might be said, we have
simply replaced one unexplained empirical fact (conditions today) with
another unexplained empirical fact (conditions in the past), so that in a
sense we are not really making progress. In our view this objection is
misguided and has roots in some unfortunate misunderstandings in the early
history of statistical mechanics.

First of all, it is perfectly reasonable to explain the present in terms of
the past. This is standard practice across the physical sciences -- from
astrophysics to geology. As a simple example, suppose that today at time $%
t_{0}$ the moon is observed to have a certain position and momentum, so that
it now occupies a particular location $(q_{0},p_{0})$ in phase space. With
the aid of Newton's laws, this fact today may be explained by the fact that
the moon was at a location $(q(t),p(t))$ in phase space at some earlier time 
$t<t_{0}$. If $t$ is very far in the past, pre-dating direct human
observation, then in practice we would deduce that the moon must have been
at $(q(t),p(t))$ at time $t$. That we have had to deduce the past from the
present would not undermine our physical intuition that the moon may be said
to be where it is now \textit{because} it was in the deduced earlier state
at an earlier time. This is normal scientific practice. On the other hand,
one can imagine the philosophical objection being raised, that the past
state is a mere deduction (or retrodiction) and not a \textit{bona fide}
`explanation' for the observed state today. It might also be suggested that
we would have a satisfactory explanation only if we could show that \textit{%
all} -- or in some sense `most' -- possible earlier states $(q(t),p(t))$ of
the moon evolve into the moon being in the state $(q_{0},p_{0})$ today.
Needless to say, most physicists would disagree with this objection (not
least because, from what we understand of lunar dynamics, such a suggestion
has no chance of being correct). The objection seems unfounded. And yet
similar objections are frequently heard in the context of statistical
mechanics. Why?

In our view the trouble stems from a mistake in the early history of the
subject. Boltzmann originally hoped to deduce the second law of
thermodynamics from mechanics alone. As is well known, this ambitious
project was fundamentally misguided. It is impossible to deduce any kind of
necessary uni-directional evolution in time in a time-reversal invariant
theory. For any initial molecular state that evolves towards thermal
equilibrium, one can always construct a time-reversed initial state that
evolves away from thermal equilibrium. Boltzmann's program was dogged by
such `reversibility objections', resulting in heated debate about the
foundations of the subject. The debate rages even today.\footnote{%
For an even-handed and scholarly review see Uffink (2007).} It is now widely
accepted that the laws of mechanics alone do not suffice: one must also
assume something about the initial conditions (such as an absence of
fine-grained microstructure, or an absence of correlations among molecular
velocities). Debates then continue about the status of the assumption about
the initial conditions, with many authors attempting to justify the
assumption on the basis of some fundamental principle or other. Running like
a thread through these debates is the expectation that, in order for the
program to succeed, it must be shown either that the required initial
conditions are `almost always' satisfied or that they are required by some
principle (where such attempts invariably lead to further controversy). In
our view, this expectation is misguided and reflects the historical error in
Boltzmann's original program. There was never any reason to expect all
allowed initial conditions to give rise to thermal relaxation, and
subsequent attempts to show that `most' initial conditions will do so, or
that the required initial conditions are consequences of some fundamental
principle or other, in effect propagate the original error (albeit in a
reduced or weaker form).

We advocate a more modest -- and in our view more reasonable -- `empirical'
approach to statistical mechanics (Valentini 1996, 2001). On this view the
observed thermodynamical behaviour is an empirical fact which must be
explained (with the aid of dynamical laws) in terms of past conditions,
where the latter are themselves also empirical. The past conditions do not
need to be `almost always' or `typically' true, nor do they need to be true
by virtue of some deep principle or other: they simply need to explain or be
consistent with what is observed. Just as in the case of the moon, where we
try to deduce -- or if necessary guess -- its state in the past given its
state today, in the case of a box of gas that is evolving towards thermal
equilibrium we try to deduce or guess the required character of the initial
(microscopic) state. For a gas, of course, there will be a set or class of
microstates yielding the observed behaviour. Unlike for the moon, we make no
attempt to deduce the exact initial micro-state. And with so many variables
involved, it is convenient to apply statistical methods. The essential aim
and method of statistical mechanics is then this. First, to find a class of
initial conditions that yields the observed behaviour. And second, to
understand the evolution of those initial conditions towards equilibrium in
terms of a general mechanism -- without having to solve the exact equations
of motion for the huge number of variables involved.

In the case of pilot-wave theory we wish to explain the observed validity of
the Born rule (\ref{Born}) for laboratory systems -- to within a certain
experimental accuracy. For example, if we prepare a large number $N$ of
hydrogen atoms in the ground state with wave function $\psi_{100}(\mathbf{x})
$, and if we measure the electron position $\mathbf{x}$ (relative to the
nucleus) for each atom, then for large $N$ we find an empirical distribution 
$\rho(\mathbf{x})$ of the schematic form%
\begin{equation}
\rho=|\psi_{100}|^{2}\pm\epsilon\ ,
\end{equation}
where $\epsilon$ characterises the accuracy to which the Born rule has been
confirmed (where this will depend on the accuracy of the position
measurements as well as on the value of $N$). How can this be explained in
terms of past conditions?

The first thing to note is that, when we encounter a hydrogen atom in the
laboratory, the atom has not been floating freely in a vacuum for billions
of years prior to us experimenting with it. The atom has a past history
during which it has interacted with other things. That past history traces
back ultimately to the formation of the earth, the solar system, our galaxy,
and ultimately merges with the history of the universe as a whole, which as
we know began with a hot and violent phase called the big bang. In fact,
every system we have access to in the laboratory has a long and violent
astrophysical history. Therefore, when attempting to explain the Born rule
today by conditions in the past, we should make use of our knowledge of that
history. In other words, when we attempt to deduce what earlier conditions
are required to explain what we observe in the laboratory today, we should
take into account what we already know about the history of the systems in
question.

The second thing to note is that the Born rule for a sub-system such as an
atom is a simple consequence of the Born rule applied to a larger system
from which the atom may have been extracted. If we consider an ensemble of
many-body systems, all with the same wave function $\Psi(q,t_{0})$ and with
an ensemble distribution $P(q,t_{0})=\left\vert \Psi(q,t_{0})\right\vert ^{2}
$ of configurations $q$ at some time $t_{0}$, then it is readily shown that
if an ensemble of sub-systems with configurations $x$ are extracted from the
parent ensemble and prepared with an effective (or reduced) wave function $%
\psi(x,t)$ at $t>t_{0}$, then the distribution of extracted configurations $x
$ will be $\rho(x,t)=\left\vert \psi(x,t)\right\vert ^{2}$ (Valentini 1991a).%
\footnote{%
A similar result was obtained by D\"{u}rr, Goldstein and Zangh\`{\i} (1992).}
In other words, equilibrium for a many-body system implies equilibrium for
extracted sub-systems (a property which is sometimes called `nesting'). This
means that we can explain the Born rule for extracted sub-systems such as
atoms if we are able to explain the Born rule for larger parent systems.

Wherever we look, in fact, we find the Born rule -- not only in the
laboratory today but also further afield. For example, the relative
intensities of atomic spectral lines emitted from the outskirts of distant
quasars agree with the Born rule (as applied to atomic transitions). The
observed cosmological helium abundance agrees with calculations based on the
Born rule (as applied to nuclear reactions in the early hot universe).
Perhaps the ultimate test of the Born rule is currently taking place in
satellite observations of the small temperature and polarization
anisotropies in the cosmic microwave background, which were caused by
classical inhomogeneities which existed at the time of photon decoupling
(around 400,000 years after the big bang), which in turn grew from classical
inhomogeneities which existed in the very early universe, and which
according to inflationary cosmology were in turn formed from quantum vacuum
fluctuations in an `inflaton' scalar field. Ultimately, on our current
understanding, the `primordial power spectrum' (the spectrum of very early
classical inhomogeneities) was generated by a Born-rule spectrum of
primordial quantum field fluctuations.

To explain the success of the Born rule, then, we can consider the earliest
possible conditions in the history of our universe. Clearly, the initial
conditions must be such as to evolve into or imply the Born rule at relevant
later times. What initial conditions should we assume?

One possibility, of course, is to simply assume that the universe began in a
state of quantum equilibrium. Below we argue that this is, in effect, the
assumption made by D\"{u}rr, Goldstein and Zangh\`{\i} (1992). Unlike in the
thermal case, we have not observed relaxation to the Born rule actually
taking place over time. All we see is the equilibrium Born rule. Since the
equations of motion preserve the Born rule over time, a simple way to
explain our observation of the Born rule now is to assume that the Born rule
was true at the beginning.

But initial equilibrium is only one possibility among uncountably many.
Given the known violent history of our universe and of everything in it, and
given the results for quantum relaxation summarised in the next section,
there clearly exists a large class of initial nonequilibrium states that
will evolve to yield the Born rule today to an excellent approximation (in
particular at the short wavelengths relevant to local physics). As we shall
see, the said initial nonequilibrium distributions may be broadly
characterised as having no fine-grained microstructure (with respect to some
coarse-graining length) while the initial wave functions are superpositions
of at least a few energy eigenstates (in order to guarantee a sufficiently
complex de Broglie velocity field). A coarse-graining $H$-theorem provides a
general mechanism in terms of which we can understand how equilibrium is
approached, without having to solve the exact equations of motion for the
system. By itself, of course, the $H$-theorem does not prove that
equilibrium is actually reached. The rate and extent of relaxation depend on
the system and on its initial wave function, as shown by extensive numerical
simulations. The Born rule today may then be understood to have arisen
dynamically, by a process of relaxation from an earlier nonequilibrium state
for which the Born rule was not valid.

Once again, in a time-reversal invariant dynamics it cannot be true that all
initial nonequilibrium states relax to equilibrium. But it does not need to
be true: non-relaxing initial conditions (with fine-grained microstructure,
or with very simple wave functions with trivial velocity fields) are ruled
out empirically, not by theoretical fiat.

As we shall see in Section 4, modern developments in theoretical and
observational cosmology make it possible to test the Born rule for quantum
fields at very early times. Thus the existence of initial equilibrium or
nonequilibrium is an empirical question -- not only in principle but also in
practice.

\section{Overview of quantum relaxation}

In this section we provide a brief overview of quantum relaxation. Further
details may be found in the cited papers.

\subsection{Coarse-graining $H$-theorem}

For a nonequilibrium ensemble of isolated systems with configurations $q$,
it follows from (\ref{Cont2}) and (\ref{Cont3}) that the ratio $%
f=\rho/\left\vert \psi\right\vert ^{2}$ is preserved along trajectories: $%
df/dt=0$. (This is the analogue of Liouville's theorem, $d\rho_{\mathrm{cl}%
}/dt=0$, for a classical phase-space density $\rho_{\mathrm{cl}}$.) The
exact $H$-function $H(t)=\int dq\ \rho\ln(\rho/|\psi|^{2})$ is then
constant, $dH/dt=0$, and there is no fine-grained relaxation. However, if we
average the densities $\rho$ and $\left\vert \psi\right\vert ^{2}$ over
small coarse-graining cells of volume $\delta V$,\footnote{%
For systems with $N$ degrees of freedom, $q=(q_{1},q_{2},...,q_{N})$ and $%
\delta V=(\delta q)^{N}$.} we may assume the absence of fine-grained
microstructure at $t=0$,\footnote{%
This will hold to arbitrary accuracy as $\delta V\rightarrow0$ if $\rho_{0}$
and $|\psi_{0}|^{2}$ are smooth functions.}%
\begin{equation}
\bar{\rho}(q,0)=\rho(q,0)\ ,\ \ \ \overline{|\psi(q,0)|^{2}}=|\psi
(q,0)|^{2}\ ,   \label{no micro}
\end{equation}
and consider the time evolution of the coarse-grained $H$-function (\ref%
{Hbar}). Defining $\tilde{f}\equiv\bar{\rho}/\overline{|\psi|^{2}}$,
straightforward manipulations show that%
\begin{equation}
\bar{H}_{0}-\bar{H}=\int dq\ |\psi|^{2}\left( f\ln(f/\tilde{f})-f+\tilde {f}%
\right)   \label{eq13}
\end{equation}
(where the subscript $0$ denotes a quantity at $t=0$ and the absence of a
subscript denotes a quantity at a general time $t$). Use of the inequality $%
x\ln(x/y)-x+y\geq0$ -- for all real and non-negative $x$, $y$, with equality
if and only if $x=y$ -- then implies the coarse-graining $H$-theorem
(Valentini 1991a, 1992)\footnote{%
As is well known for the classical case, the result (\ref{Hthm_qu}) is
time-symmetric, with $t=0$ a local maximum of $\bar{H}(t)$.}%
\begin{equation}
\bar{H}(t)\leq\bar{H}(0)\ .   \label{Hthm_qu}
\end{equation}
From (\ref{eq13}) it also follows that (\ref{Hthm_qu}) becomes a strict
inequality, $\bar{H}(t)<\bar{H}(0)$, when $f\neq\tilde{f}$. Since $%
\left\vert \psi\right\vert ^{2}$ remains smooth this occurs when $\rho\neq%
\bar{\rho}$, that is, when $\rho$ develops fine-grained structure -- which
it generally will for non-trivial velocity fields that vary over the
coarse-graining cells.

The quantity $\bar{H}$ is equal to minus the relative entropy of $\bar{\rho}$
with respect to $\overline{|\psi |^{2}}$. As already noted, $\bar{H}$ is
bounded from below by zero and the minimum $\bar{H}=0$ is attained if and
only if $\bar{\rho}=\overline{|\psi |^{2}}$. Thus a decrease of $\bar{H}$
quantifies relaxation to the Born rule.\footnote{%
The quantity $\bar{H}$ is also equal to the well-known (in mathematical
statistics) Kullback-Leibler divergence $D_{\mathrm{KL}}(\bar{\rho}\parallel 
\overline{|\psi |^{2}})$, which measures how $\bar{\rho}$ differs from $%
\overline{|\psi |^{2}}$.}

The result (\ref{Hthm_qu}) formalises an intuitive understanding of
relaxation in terms of a mixing of two `fluids' with densities $\rho$ and $%
|\psi|^{2}$ in configuration space. These obey the same continuity equation
and are therefore `stirred' by the same velocity field $\dot{q}$. For a
sufficiently complicated flow, $\rho$ and $|\psi|^{2}$ tend to become
indistinguishable on a coarse-grained level (Valentini 1991a). This is
similar to the classical stirring of two fluids that was famously discussed
by Gibbs (1902), where his fluids were analogous to the classical
phase-space densities $\rho _{\mathrm{cl}}$ (for a general ensemble) and $%
\rho_{\mathrm{eq}}=\mathrm{const.}$ (for an equilibrium ensemble) on the
energy surface, and where the mixing of $\rho_{\mathrm{cl}}$ and $\rho_{%
\mathrm{eq}}$ may be quantified by a decrease of the classical $H$-function $%
\bar{H}_{\mathrm{cl}}=\int dq\ \bar{\rho}_{\mathrm{cl}}\ln\left( \bar{\rho}_{%
\mathrm{cl}}/\overline{\rho_{\mathrm{eq}}}\right) $. In both cases the
nonequilibrium density develops fine-grained structure while the equilibrium
density remains smooth. The increase of the `subquantum entropy' $\bar{S}=-%
\bar{H}$ may be associated with the mixing of $\rho$ and $|\psi|^{2}$ in
configuration space, just as the increase of the Gibbs entropy $\bar{S}_{%
\mathrm{Gibbs}}=-\bar {H}_{\mathrm{cl}}$ may be associated with the mixing
of $\rho_{\mathrm{cl}}$ and $\rho_{\mathrm{eq}}$ in phase space.

Note that we use the word `mixing' informally in the simple sense of
`stirring' (as in Gibbs' original analogy). Mathematical mixing is defined
by an infinite-time limit.\footnote{%
Formally, a dynamical system (with measure $\mu $ on a set $\Gamma $ subject
to measure-preserving transformations $T_{t}$) is `mixing' if and only if $%
\lim_{t\rightarrow \infty }\mu (T_{t}A\cap B)=\mu (A)\mu (B)$ for all
relevant\ subsets $A$ and $B$ of $\Gamma $.} Its physical relevance is
therefore questionable, and in any case it might not apply to realistic
systems even in that limit (Uffink 2007). In our view the above process --
whereby $\rho $ develops fine-grained structure while $|\psi |^{2}$ does not
-- is the essential physical mechanism that drives quantum relaxation over
realistic timescales. Though as noted, the actual rate and extent of
relaxation will depend on the system.

An assumption about initial conditions is of course necessary to explain
relaxation in a time-reversal invariant theory. We offer no `principle' to
justify the initial conditions (\ref{no micro}). In the spirit of our
empirical approach, they are justified only by the extent to which they help
us explain observations. We take (\ref{no micro}) to be matters of fact
about our world, while acknowledging that in principle they could be false.
Their truth or falsity is ultimately a matter for experiment.

It is also worth remarking that our approach (like its classical Gibbsean
counterpart) does not rely on any particular interpretation of probability
theory. The density $\rho$ might represent a subjective probability for a
single system, a distribution over a theoretical ensemble, or the
distribution of a real ensemble of existing systems, according to taste or
requirement.

Note also that, for a finite real ensemble of $N$ systems, the actual
density $\rho$ will be a sum of delta-functions, which can approach a smooth
function only in the large-$N$ limit (Valentini 1992, pp. 18, 36). To obtain
a density with no fine-grained structure on a coarse-graining scale $\delta V
$, we must of course take the appropriate large-$N$ limit \textit{before}
considering small $\delta V$.\footnote{%
This resolves a concern raised by Norsen (2018, p. 16) that the condition (%
\ref{no micro}) on $\rho_{0}$ cannot be satisfied for realistic finite
ensembles. The same concern arises, of course, in the classical Gibbsean
approach and has the same resolution.}

\subsection{Numerical simulations}

Extensive numerical simulations demonstrate that quantum relaxation takes
place efficiently for wave functions $\psi$ that are superpositions of
multiple energy eigenstates (Valentini and Westman 2005; Towler, Russell and
Valentini 2012; Colin 2012; Abraham, Colin and Valentini 2014). An example
is shown in Figure 1, for a two-dimensional oscillator in a superposition of 
$M=25$ modes and with an initial Gaussian nonequilibrium density $\rho_{0}$.
The top row displays the time evolution of the (coarse-grained) equilibrium
density $\left\vert \psi\right\vert ^{2}$, while the bottom row displays
relaxation of the actual (coarse-grained) density $\rho$. Comparable results
are obtained for superpositions with as little as $M=4$ modes, and also for
a two-dimensional box.

\begin{figure}[tbp]
\begin{center}
\includegraphics[width=0.8\textwidth]{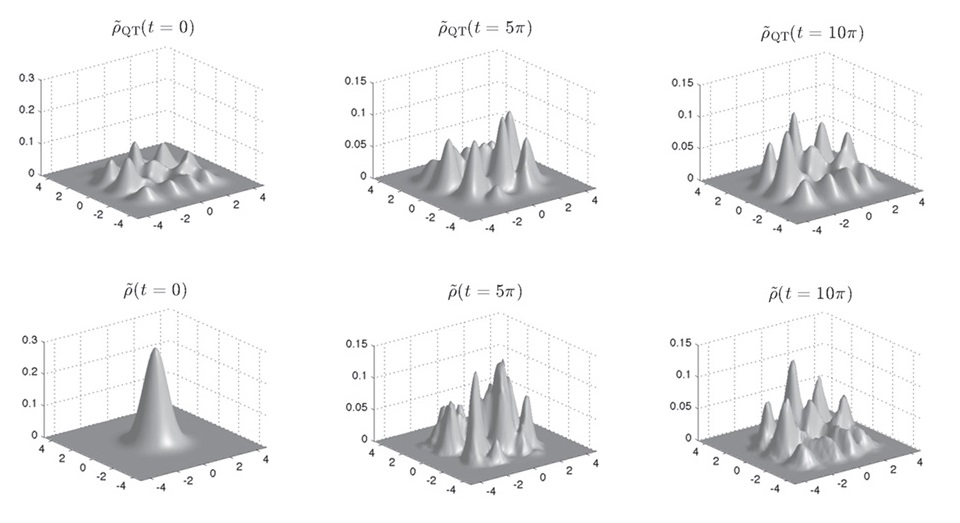}
\end{center}
\caption{Illustration of numerical quantum relaxation for an oscillator in a
superposition of $M=25$ modes (Abraham, Colin and Valentini 2014). The wave
function has period $2\protect\pi$. The top row shows the evolving
coarse-grained density $\tilde{\protect\rho}_{\mathrm{QT}}$ as predicted by
quantum theory, while the bottom row shows the coarse-grained relaxation of
an initial nonequilibrium density $\protect\rho(x,y,0)=(1/\protect\pi%
)e^{-(x^{2}+y^{2})}$ (where tildes denote a smoothed coarse-graining with
overlapping cells). After five periods the coarse-grained densities are
almost indistinguishable. (Note the different vertical scale at $t=0$.)}
\end{figure}

In all of these simulations $\bar{H}$ is found to decay approximately
exponentially with time: $\bar{H}(t)\approx\bar{H}_{0}e^{-t/\tau}$ for some
constant $\tau$ whose value depends on the initial wave function as well as
on the coarse-graining length (Towler, Russell and Valentini 2012). An
example is shown in Figure 2.

\begin{figure}[tbp]
\begin{center}
\includegraphics[width=0.8\textwidth]
{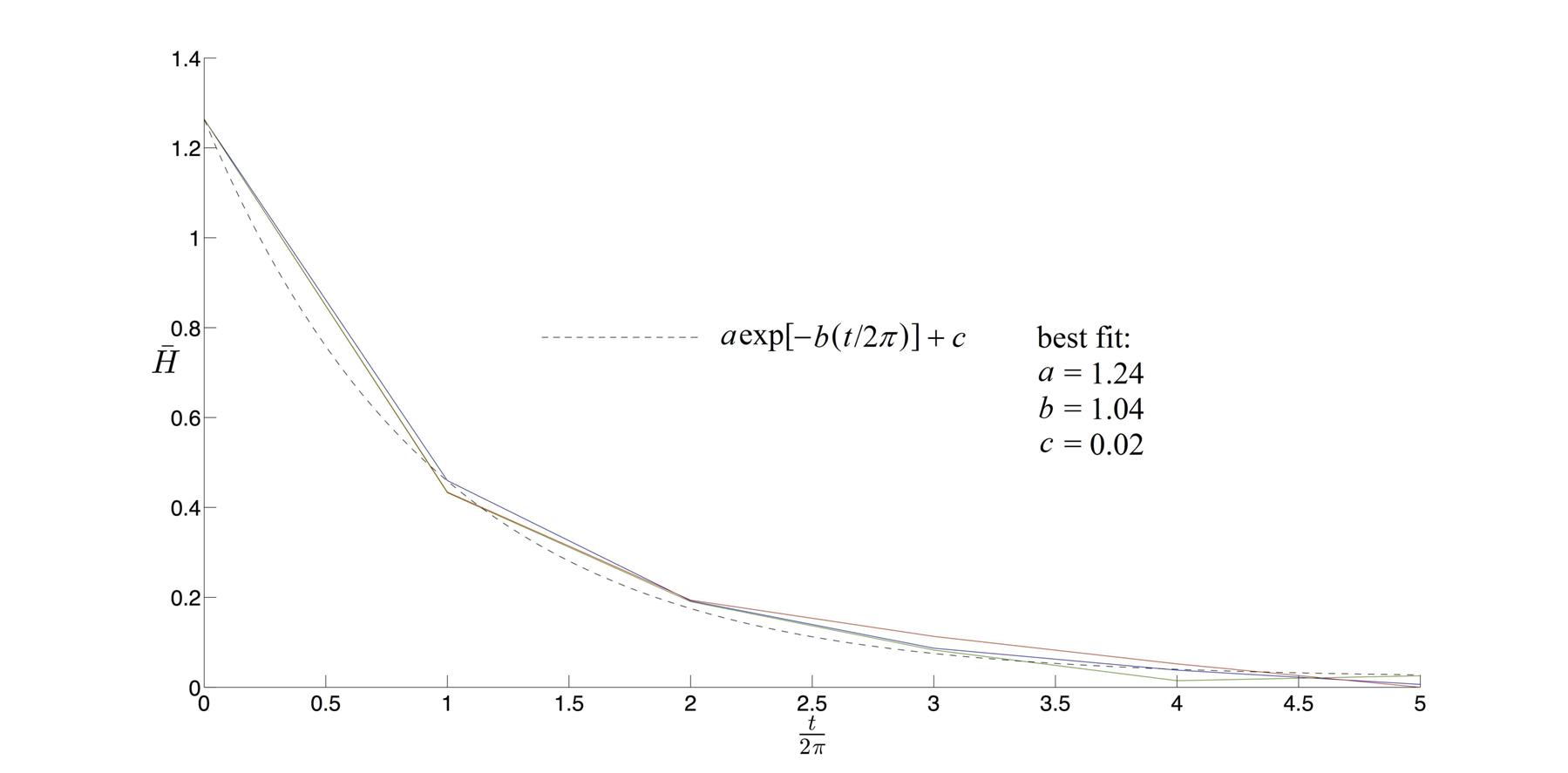}
\end{center}
\caption{Approximate exponential decay of $\bar{H}(t)$ for the same
simulation displayed in Figure 1 (Abraham, Colin and Valentini 2014). The
error in $\bar{H}$ is estimated by running three separate simulations with
different numerical grids (the solid curves). Fitting to an exponential
(dashed curve) yields a best-fit residue $c=0.02$ that is comparable to the
late-time error, indicating no discernible late-time residue in $\bar{H}$.
We find a decay timescale $\protect\tau=2\protect\pi/b\simeq\allowbreak6$
(units $\hbar=m=\protect\omega=1$).}
\end{figure}

Some simulations show a small but discernible non-zero `residue' in $\bar{H}$
at large times (unlike the case displayed in Figure 2), indicating that
equilibrium is not reached exactly (Abraham, Colin and Valentini 2014). For
these cases, the trajectories tend to show some degree of confinement (not
fully exploring the support of $\left\vert \psi\right\vert ^{2}$). Numerical
evidence shows that this is less likely to happen for larger $M$. Because
all laboratory systems have a long and violent astrophysical history, during
which the relevant value of $M$ will have been very large, we may expect
that in the remote past they will have reached equilibrium on a very small
coarse-graining scale.

\subsection{The early universe}

This motivates us to consider quantum relaxation in the early universe. This
may be discussed for a free massless scalar field $\phi$ on flat expanding
space with spacetime metric%
\begin{equation}
d\tau^{2}=dt^{2}-a^{2}d\mathbf{x}^{2}   \label{metric}
\end{equation}
and a scale factor $a(t)\propto t^{1/2}$ (corresponding to a
radiation-dominated expansion), where $t$ is standard cosmological time and
physical wavelengths $\lambda_{\mathrm{phys}}(t)$ are proportional to $a(t)$%
. The Fourier components $\phi_{\mathbf{k}}(t)$ may be written as%
\begin{equation}
\phi_{\mathbf{k}}=\frac{\sqrt{V}}{(2\pi)^{3/2}}\left( q_{\mathbf{k}1}+iq_{%
\mathbf{k}2}\right) \ ,
\end{equation}
where $q_{\mathbf{k}1}$, $q_{\mathbf{k}2}$ are real and $V$ is a
normalisation volume. The field Hamiltonian then takes the form $\hat{H}%
=\sum_{\mathbf{k}r}\hat{H}_{\mathbf{k}r}$, where $\hat{H}_{\mathbf{k}r}$ ($%
r=1,2$) coincides with the Hamiltonian of a harmonic oscillator of mass $%
m=a^{3}$ and angular frequency $\omega=k/a$ (Valentini 2007, 2008a, 2010a).
Thus a single (unentangled) field mode $\mathbf{k}$ in the early universe is
mathematically equivalent to a two-dimensional oscillator with a
time-dependent mass. This system is in turn equivalent to an ordinary
oscillator (with constant mass and angular frequency) but with $t$ replaced
by a `retarded time' $t_{\mathrm{ret}}=t_{\mathrm{ret}}(t,k)$ that depends
on $k$ (Colin and Valentini 2013). It is found that quantum relaxation
depends crucially on how $\lambda_{\mathrm{phys}}$ compares with the Hubble
radius $H^{-1}=a/\dot{a}$. For short wavelengths $\lambda_{\mathrm{phys}%
}<<H^{-1}$ we find $t_{\mathrm{ret}}(t,k)\rightarrow t$ and we recover
physics on static flat space, with the same rapid relaxation illustrated in
Figure 1 for the oscillator. Whereas for long wavelengths $\lambda_{\mathrm{%
phys}}\gtrsim H^{-1}$ we find $t_{\mathrm{ret}}(t,k)<t$ and relaxation is
retarded or suppressed (Valentini 2008a; Colin and Valentini 2013).

The suppression of quantum relaxation at long (super-Hubble) wavelengths is
illustrated in Figure 3, where we show the time evolution of a
nonequilibrium field mode with $\lambda_{\mathrm{phys}}=10H^{-1}$ (at
initial time $t_{i}$). Over the given time interval $(t_{i},t_{f})$
relaxation proceeds but is incomplete: in particular, the final
nonequilibrium width (or variance) is smaller than the final equilibrium
width. In contrast, in Figure 4 we show relaxation with the same initial
conditions and over the same time interval $(t_{i},t_{f})$ but with no
spatial expansion: relaxation now takes place essentially completely and the
final widths match closely.

\begin{figure}[tbp]
\begin{center}
\includegraphics[width=0.8\textwidth]{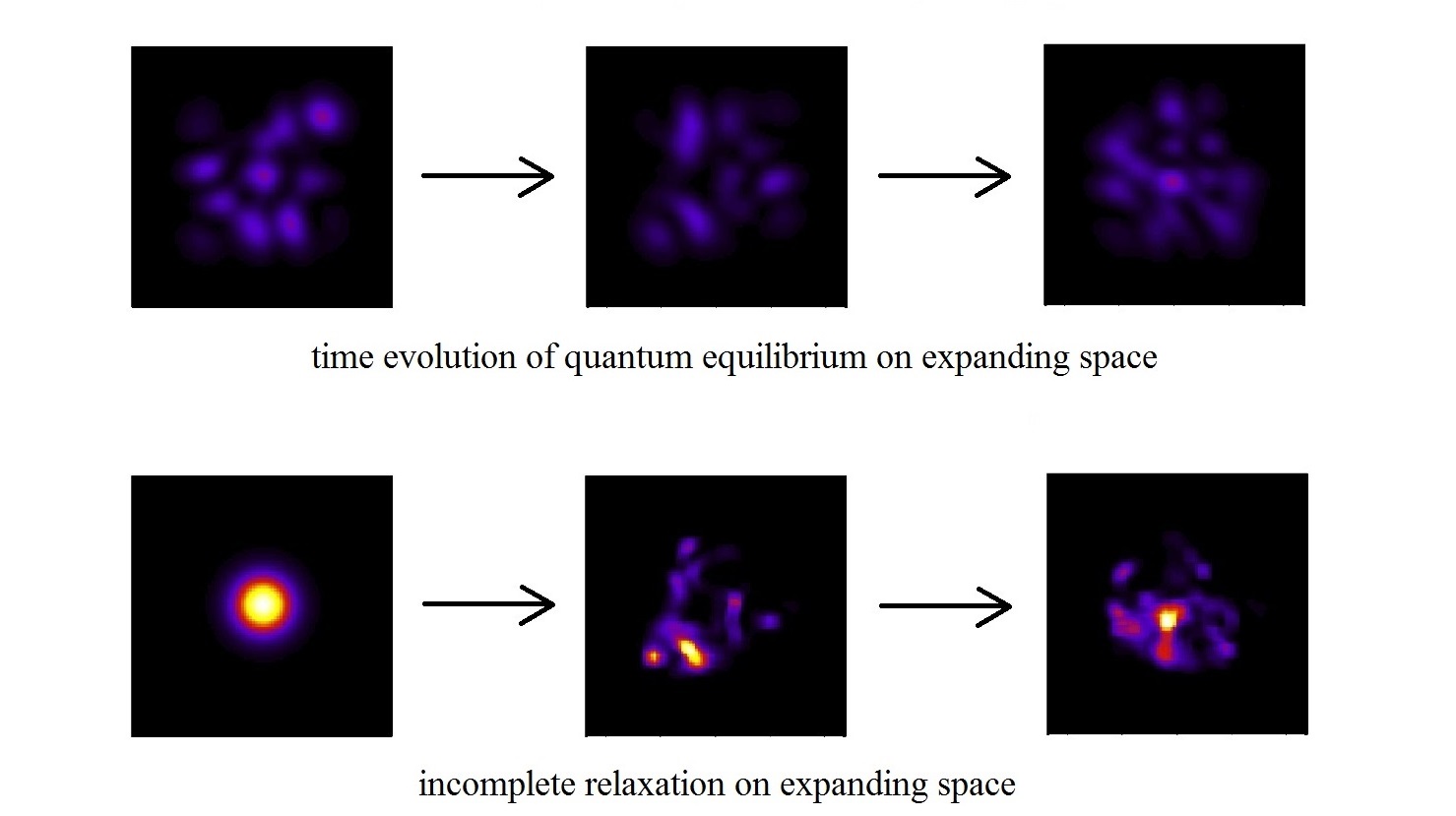}
\end{center}
\caption{Incomplete quantum relaxation for a super-Hubble field mode $%
\protect\phi_{\mathbf{k}}$ on expanding space over a time interval $%
(t_{i},t_{f})$ (Colin and Valentini 2013). The final nonequilibrium width is
noticeably smaller than the final equilibrium width.}
\end{figure}

\begin{figure}[tbp]
\begin{center}
\includegraphics[width=0.8\textwidth]{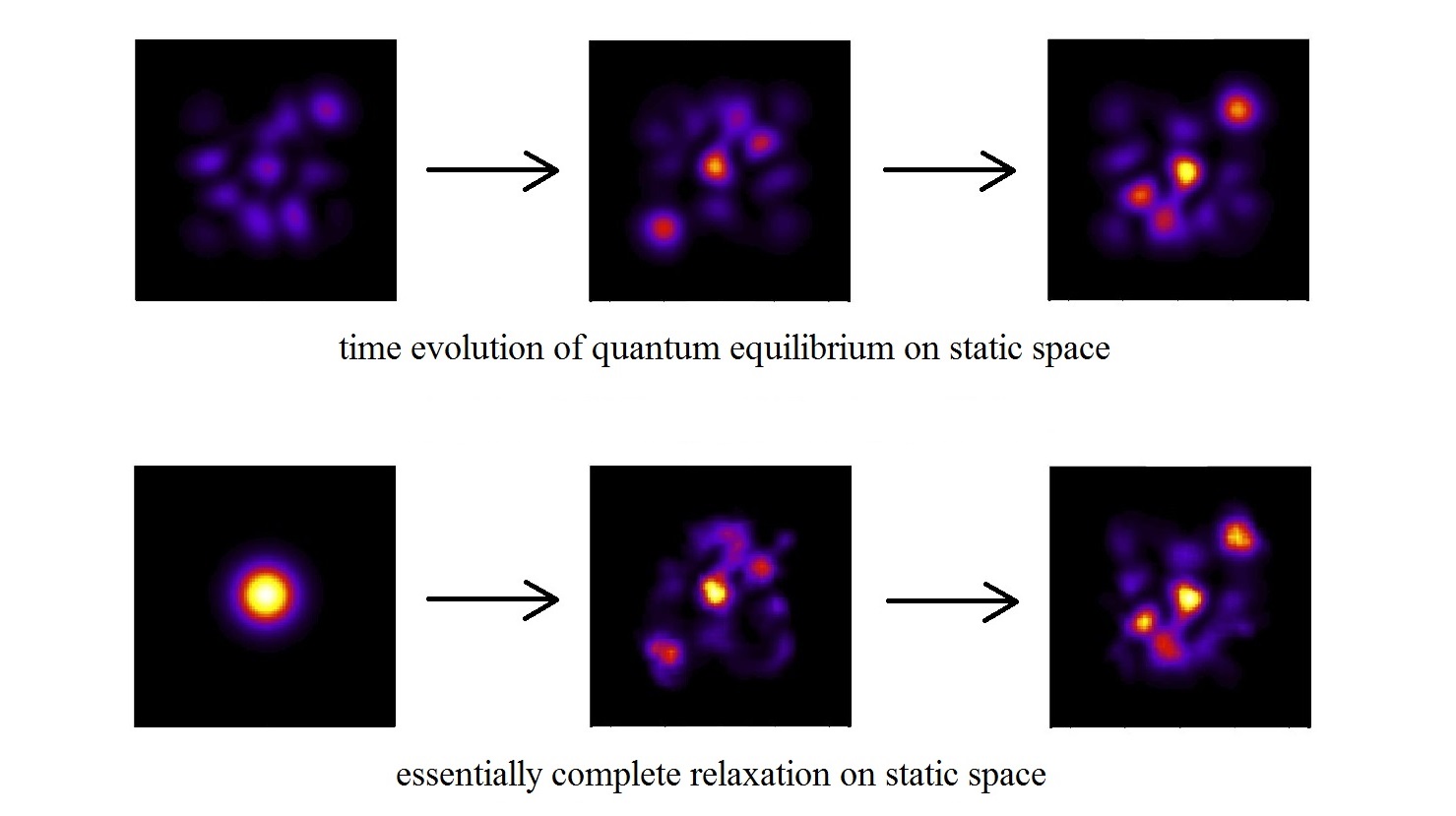}
\end{center}
\caption{Essentially complete quantum relaxation for a field mode $\protect%
\phi_{\mathbf{k}}$ on static space over the same time interval $(t_{i},t_{f})
$ as in Figure 3 and with the same initial conditions (Colin and Valentini
2013). The final nonequilibrium width closely matches the final equilibrium
width.}
\end{figure}

Cosmologically speaking, the reduced width of the final distribution in
Figure 3 is of particular interest. For a mode with wavenumber $k$ we may
write%
\begin{equation}
\left\langle \left\vert \phi_{\mathbf{k}}\right\vert ^{2}\right\rangle
=\left\langle \left\vert \phi_{\mathbf{k}}\right\vert ^{2}\right\rangle _{%
\mathrm{QT}}\xi(k)\ ,   \label{var_noneq}
\end{equation}
where $\left\langle ...\right\rangle $ and $\left\langle ...\right\rangle _{%
\mathrm{QT}}$ denote respective nonequilibrium and equilibrium expectation
values. The function $\xi(k)$ quantifies the degree of primordial quantum
nonequilibrium as a function of $k$. For the case shown in Figure 3, at the
final time we clearly have $\xi(k)<1$ (corresponding to a `power deficit').
As a general trend we expect $\xi(k)$ to be smaller for smaller $k$, where
longer wavelengths imply less relaxation. This has been broadly confirmed by
running extensive simulations for varying values of $k$ and plotting the
function $\xi=\xi(k)$. The resulting curves show oscillations of magnitude $%
\lesssim10\%$. As a first approximation we may ignore these, in which case
we find a good fit to the function%
\begin{equation}
\xi(k)=\tan^{-1}(c_{1}\frac{k}{\pi}+c_{2})-\frac{\pi}{2}+c_{3}\ , 
\label{atan}
\end{equation}
where the parameters $c_{1}$, $c_{2}$ and $c_{3}$ depend on the initial
state and on the time interval (Colin and Valentini 2015). This function
tends to a maximum $\xi\rightarrow c_{3}$ for $k\rightarrow\infty$, and
decreases smoothly for smaller $k$. Simulations for a range of different
initial nonequilibria -- all assumed to have a narrower-than-quantum initial
width\footnote{%
Heuristically, it seems natural to assume initial conditions with a
subquantum statistical spread (so the initial state contains less noise than
a conventional quantum state), but in principle this assumption could of
course be incorrect.} -- show a good fit to the curve (\ref{atan}) (ignoring
the oscillations) (Colin and Valentini 2016). Thus, with some mild
assumptions about the initial state, the `deficit function' (\ref{atan}) is
a robust approximate prediction of the cosmological quantum relaxation
scenario (for a free scalar field at the end of a radiation-dominated era).

\section{Testing the primordial Born rule with cosmological data}

We now outline how the Born rule may be tested at very early times with
cosmological data.

The temperature of the cosmic microwave background (CMB) today is slightly
anisotropic. It is customary to write a spherical harmonic expansion%
\begin{equation}
\frac{\Delta T(\mathbf{\hat{n}})}{\bar{T}}=\sum_{l=2}^{\infty}%
\sum_{m=-l}^{+l}a_{lm}Y_{lm}(\mathbf{\hat{n}})   \label{sph_har}
\end{equation}
for the measured anisotropy $\Delta T(\mathbf{\hat{n}})\equiv T(\mathbf{%
\hat {n}})-\bar{T}$, where the unit vector $\mathbf{\hat{n}}$ labels points
on the sky and $\bar{T}$ is the mean temperature. As noted the CMB was
formed around 400,000 years after the big bang, and its small anisotropies
reflect small inhomogeneities of the universe at that time. Thus the
coefficients $a_{lm}$ are generated by the Fourier-space `primordial
curvature perturbation' $\mathcal{R}_{\mathbf{k}}$ according to the formula
(Lyth and Riotto 1999)%
\begin{equation}
a_{lm}=\frac{i^{l}}{2\pi^{2}}\int d^{3}\mathbf{k}\ \mathcal{T}(k,l)\mathcal{R%
}_{\mathbf{k}}Y_{lm}(\mathbf{\hat{k}})\ ,   \label{alm}
\end{equation}
where the `transfer function' $\mathcal{T}(k,l)$ encodes the relevant
astrophysics.

Cosmologists generally assume that the measured function $\Delta T(\mathbf{%
\hat{n}})$ is a single realisation of a random variable with a probability
distribution $P[\Delta T(\mathbf{\hat{n}})]$ associated with a `theoretical
ensemble' (which may be interpreted according to taste). It is usual to
assume `statistical isotropy', which means that $P$ is invariant under a
rotation $\mathbf{\hat{n}}\rightarrow\mathbf{\hat{n}}^{\prime}$ (that is, $%
P[\Delta T(\mathbf{\hat{n}}^{\prime})]=P[\Delta T(\mathbf{\hat{n}})]$). This
implies that the ensemble average $\left\langle \Delta T(\mathbf{\hat{n}}%
_{1})\Delta T(\mathbf{\hat{n}}_{2})\right\rangle $ depends only on the angle
between $\mathbf{\hat{n}}_{1}$ and $\mathbf{\hat{n}}_{2}$, which in turn
implies (Hajian and Souradeep 2005, appendix B)%
\begin{equation}
\left\langle a_{l^{\prime}m^{\prime}}^{\ast}a_{lm}\right\rangle =\delta
_{ll^{\prime}}\delta_{mm^{\prime}}C_{l}\ ,   \label{iso}
\end{equation}
where%
\begin{equation}
C_{l}\equiv\left\langle \left\vert a_{lm}\right\vert ^{2}\right\rangle
\end{equation}
(the `angular power spectrum') is independent of $m$. Thus, while for fixed $%
l$ there are $2l+1$ different quantities $\left\vert a_{lm}\right\vert ^{2}$%
, statistical isotropy implies that they each have the same ensemble mean $%
C_{l}$.

For our one observed sky we may define a measured mean statistic%
\begin{equation*}
C_{l}^{\mathrm{sky}}\equiv\frac{1}{2l+1}\sum_{m=-l}^{+l}\left\vert
a_{lm}\right\vert ^{2}\ . 
\end{equation*}
This obviously satisfies $\left\langle C_{l}^{\mathrm{sky}}\right\rangle
=C_{l}$. Thus $C_{l}^{\mathrm{sky}}$ (measured for one sky) is an unbiased
estimate of $C_{l}$ for the ensemble. Furthermore, assuming a Gaussian
distribution it may be shown that $C_{l}^{\mathrm{sky}}$ has a `cosmic
variance'%
\begin{equation}
\frac{\Delta C_{l}^{\mathrm{sky}}}{C_{l}}=\sqrt{\frac{2}{2l+1}}\ . 
\label{CV}
\end{equation}
Thus for large $l$ we expect to find $C_{l}^{\mathrm{sky}}\approx C_{l}$,
whereas for small $l$ the accuracy is limited.

It is also usually assumed that the theoretical ensemble for $\mathcal{R}$
is statistically homogeneous (that is, the probability distribution $P[%
\mathcal{R}(\mathbf{x})]$ is invariant under spatial translations). This
implies that $\left\langle \mathcal{R}(\mathbf{x})\mathcal{R}(\mathbf{x}%
^{\prime})\right\rangle $ depends only on $\mathbf{x-x}^{\prime}$, which
implies%
\begin{equation}
\left\langle \mathcal{R}_{\mathbf{k%
{\acute{}}%
}}^{\ast}\mathcal{R}_{\mathbf{k}}\right\rangle =\delta_{\mathbf{kk}%
{\acute{}}%
}\left\langle \left\vert \mathcal{R}_{\mathbf{k}}\right\vert
^{2}\right\rangle \ .   \label{RkRk}
\end{equation}
From (\ref{alm}) and (\ref{RkRk}) it follows that%
\begin{equation}
C_{l}=\frac{1}{2\pi^{2}}\int_{0}^{\infty}\frac{dk}{k}\ \mathcal{T}^{2}(k,l)%
\mathcal{P}_{\mathcal{R}}(k)\ ,   \label{Cl2}
\end{equation}
where%
\begin{equation}
\mathcal{P}_{\mathcal{R}}(k)\equiv\frac{4\pi k^{3}}{V}\left\langle
\left\vert \mathcal{R}_{\mathbf{k}}\right\vert ^{2}\right\rangle 
\label{PPS}
\end{equation}
is the `primordial power spectrum' for $\mathcal{R}_{\mathbf{k}}$. Thus
measurements for a single sky constrain the spectrum $\mathcal{P}_{\mathcal{R%
}}(k)$ -- which is a property of the theoretical ensemble.

Note that in the discussion so far the quantities $a_{lm}$, $C_{l}$ and $%
\mathcal{R}_{\mathbf{k}}$ are treated classically.\footnote{%
Goldstein, Struyve and Tumulka (2015) give a rather confused account of the
relation between primordial perturbations and CMB anisotropies, missing in
particular the crucial role played by statistical isotropy.}

What is the origin of the spectrum (\ref{PPS})? According to inflationary
cosmology, during a very early period of approximately exponential expansion
a perturbation $\phi_{\mathbf{k}}$ of the `inflaton field' generates a
curvature perturbation $\mathcal{R}_{\mathbf{k}}\propto\phi_{\mathbf{k}}$
(once the physical wavelength of the mode exceeds the Hubble radius) (Liddle
and Lyth 2000). The quantum-theoretical variance $\left\langle |\phi_{%
\mathbf{k}}|^{2}\right\rangle _{\mathrm{QT}}$ is calculated from quantum
field theory assuming the Born rule for an appropriate vacuum state. From
this we readily obtain the corresponding variance $\left\langle |\mathcal{R}%
_{\mathbf{k}}|^{2}\right\rangle _{\mathrm{QT}}$ and hence a
quantum-theoretical prediction $\mathcal{P}_{\mathcal{R}}^{\mathrm{QT}}(k)$
for the spectrum of $\mathcal{R}_{\mathbf{k}}$ (approximately flat with a
slight tilt). If instead we have a nonequilibrium variance (\ref{var_noneq})
for $\phi_{\mathbf{k}}$, the predicted spectrum for $\mathcal{R}_{\mathbf{k}}
$ is corrected by the factor $\xi(k)$:%
\begin{equation}
\mathcal{P}_{\mathcal{R}}(k)=\mathcal{P}_{\mathcal{R}}^{\mathrm{QT}%
}(k)\xi(k)\ .   \label{PS_noneq}
\end{equation}
Inserting this into (\ref{Cl2}) yields a corrected angular power spectrum $%
C_{l}$. From measurements of the CMB we may then set observational bounds on 
$\xi(k)$ -- that is, on corrections to the Born rule in the very early
universe (Valentini 2010a).

To obtain a prediction for $\xi(k)$, we may for example assume that quantum
relaxation took place during a pre-inflationary era. It is not uncommon for
cosmologists to assume that such an era was radiation-dominated.\footnote{%
See, for example, Wang and Ng (2008).} In this case, at the end of
pre-inflation we expect the nonequilibrium variance to be corrected by a
deficit function of the approximate form (\ref{atan}). It may be shown that
relaxation does not take place during inflation itself (Valentini 2010a). If
we make the simplifying assumption that the spectrum is unaffected by the
transition from pre-inflation to inflation, we obtain a prediction for a
corrected primordial power spectrum (\ref{PS_noneq}) with $\xi(k)$ of the
form (\ref{atan}) (with three unknown parameters). Note, however, that if
the lengthscale $c_{1}$ is too large the dip in the spectrum will be
essentially unobservable (even if it exists).

CMB data from the \textit{Planck} satellite show hints of a power deficit at
large scales (small $k$ and low $l$) (Aghanim \textit{et al}. 2016). Because
of the large cosmic variance in this region, it is however difficult to draw
firm conclusions. Extensive data analysis shows that the predicted deficit (%
\ref{atan}) fits the data more or less as well as the standard `power-law'
model (where the evaluated significance takes into acount the larger number
of parameters) (Vitenti, Peter and Valentini 2019). This is a modest
success, in the sense that with three extra parameters the significance
could have been worse. But evidence from this fitting alone neither supports
nor rules out the prediction (\ref{atan}). If one is inclined to invoke
Ockham's razor in favour of the simplest cosmological model, then the
observed low-power anomaly may reasonably be regarded as a statistical
fluctuation. To obtain evidence for or against our model we must include
more detailed predictions, such as oscillations around the curve (\ref{atan}%
) (Colin and Valentini 2015; Kandhadai and Valentini 2019) or possible
violations of statistical isotropy (Valentini 2015).

\section{Critique of `typicality' as an explanation for the Born rule}

In this section we provide a critical assessment of the typicality approach
to understanding the Born rule.

\subsection{Typicality, probability, and intrinsic likelihood}

As we noted in the Introduction, the Bohmian mechanics school attempts to
explain the success of the Born rule today by appealing to a notion of
`typicality' for the initial configuration $q_{\mathrm{univ}}(0)$ of the
universe. In this approach, if $\Psi_{\mathrm{univ}}(q_{\mathrm{univ}},0)$
is the initial universal wave function then $|\Psi_{\mathrm{univ}}(q_{%
\mathrm{univ}},0)|^{2}$ is assumed to be the `natural measure' on the set of
initial universal configurations $q_{\mathrm{univ}}(0)$.

In our view the typicality approach amounts to assuming, without
justification, that the universe as a whole began in quantum equilibrium
(Valentini 1996, 2001). The approach then seems circular. Defenders of the
approach might attempt to avoid the charge of circularity by claiming that
typicality and probability are conceptually distinct. In our view, however,
typicality is synonomous with probability. The Bohmian mechanics school
employs the word `typicality' when referring to probability for the whole
universe and employs the word `probability' when referring to probability
for sub-systems. But in our view the two words mean the same thing.

\begin{center}
\textit{Typicality and the Born rule}
\end{center}

D\"{u}rr, Goldstein and Zangh\`{\i} (1992) showed that, if we consider
sub-systems within the universe, we will obtain the Born rule for `almost
all' initial configurations $q_{\mathrm{univ}}(0)$ -- where `almost all' is
defined with respect to $|\Psi_{\mathrm{univ}}(q_{\mathrm{univ}},0)|^{2}$.
Thus it may be said that for sub-systems the quantum equilibrium
distribution $\rho =|\psi|^{2}$ is `typical', where the notion of
`typicality' is defined with respect to the measure $|\Psi_{\mathrm{univ}%
}(q_{\mathrm{univ}},0)|^{2}$.

This result may be illustrated by a simple example. We consider a model
universe at $t=0$ containing a large number $n$ of unentangled sub-systems
all with the same initial wave function $\psi(q,0)$ and with initial
configurations $q_{1}(0),\ q_{2}(0),\ ...,\ q_{n}(0)$ that generally vary
from one sub-system to another. We may write%
\begin{equation}
q_{\mathrm{univ}}(0)=(q_{1}(0),\ q_{2}(0),\ ...,\ q_{n}(0))
\end{equation}
and%
\begin{equation}
\Psi_{\mathrm{univ}}(q_{\mathrm{univ}},0)=\psi(q_{1},0)\psi(q_{2},0)...%
\psi(q_{n},0)\ .   \label{Psi_univ_prod}
\end{equation}
For large $n$, a given initial configuration $q_{\mathrm{univ}}(0)$
determines an initial distribution $\rho(q,0)$ over the ensemble of
sub-systems. Thus we have a schematic correspondence%
\begin{equation*}
q_{\mathrm{univ}}(0)\longleftrightarrow\rho(q,0)\ \ \ \ (n\rightarrow
\infty)\ . 
\end{equation*}

Note that the induced distribution $\rho(q,0)$ need not be equal to (or even
close to) $|\psi(q,0)|^{2}$. On the other hand, if we adopt the measure $%
|\Psi_{\mathrm{univ}}(q_{\mathrm{univ}},0)|^{2}$ on the set of possible
initial configurations $q_{\mathrm{univ}}(0)$, then it is easy to see that,
with respect to this measure, in the limit $n\rightarrow\infty$ almost all
points $q_{\mathrm{univ}}(0)$ correspond to the Born-rule distribution $%
\rho(q,0)=|\psi(q,0)|^{2}$. This is because, with respect to the universal
measure%
\begin{equation*}
|\Psi_{\mathrm{univ}}(q_{\mathrm{univ}},0)|^{2}=|\psi(q_{1},0)|^{2}|\psi
(q_{2},0)|^{2}...|\psi(q_{n},0)|^{2}\ , 
\end{equation*}
in effect we have $n$ independent and identically-distributed random
variables $q_{1},\ q_{2},\ q_{3},\ ...,\ q_{n}$, each with the same
probability distribution $|\psi(q,0)|^{2}$. In the limit $n\rightarrow\infty$
we then necessarily find $\rho(q,0)=|\psi(q,0)|^{2}$.

It might then appear that nonequilibrium configurations -- that is,
configurations $q_{\mathrm{univ}}(0)$ corresponding to distributions $%
\rho(q,0)\neq|\psi(q,0)|^{2}$ -- comprise a vanishingly small set (in the
limit $n\rightarrow\infty$) and may then be regarded as intrinsically
unlikely or `untypical'. But this conclusion rests crucially on the choice
of measure $|\Psi_{\mathrm{univ}}(q_{\mathrm{univ}},0)|^{2}$. For example,
if instead we choose a measure $|\Psi_{\mathrm{univ}}(q_{\mathrm{univ}%
},0)|^{4}$ (up to overall normalisation), then by the same argument almost
all configurations $q_{\mathrm{univ}}(0)$ now correspond to a nonequilibrium
distribution $\rho(q,0)\propto|\psi(q,0)|^{4}$ for sub-systems. More
generally, if we choose a measure $|\Psi_{\mathrm{univ}}(q_{\mathrm{univ}%
},0)|^{p}$ (for constant $p>0$), we will almost always obtain a
nonequilibrium distribution $\rho(q,0)\propto|\psi(q,0)|^{p}$ for
sub-systems. The typicality approach then seems circular: by assuming a
universal Born-rule measure $|\Psi _{\mathrm{univ}}(q_{\mathrm{univ}},0)|^{2}
$, one is simply assuming the Born rule at the initial time $t=0$ (Valentini
1996, 2001).

In our illustrative example $\Psi_{\mathrm{univ}}(q_{\mathrm{univ}},0)$
takes the simple form (\ref{Psi_univ_prod}) and we only consider
measurements at $t=0$. The original argument by D\"{u}rr, Goldstein and Zangh%
\`{\i} (1992) is more general than this and includes a discussion of time
ensembles of measurements. But the key objection remains: the Born rule is
guaranteed to hold for sub-systems in the early universe only because the
Born rule is assumed to hold for the whole universe at $t=0$.

It is important to emphasise that there is a qualitative difference between
cases with large-but-finite $n$ and the literal limit $n\rightarrow \infty $%
. For any finite $n$, however large, a set of points $q_{\mathrm{univ}%
}=(q_{1},\ q_{2},\ ...,\ q_{n})$ that has zero measure with respect to $%
|\Psi _{\mathrm{univ}}(q_{\mathrm{univ}},0)|^{2}$ will arguably have zero
measure with respect to any reasonable density function on configuration
space. Thus, for example, if a particle moving in two spatial dimensions has
an initial wave function $\psi (x,y,0)$ then points and lines in the
two-dimensional configuration space will have zero measure with respect to $%
|\psi (x,y,0)|^{2}$. Those same points and lines will also have zero
Lebesgue measure (or zero area), and furthermore they will have zero measure
with respect to any density proportional to $|\psi (x,y,0)|^{p}$ (with $p>0$%
). Much the same may be said for general configuration spaces. This means
that, for finite $n$, if a set of points $q_{\mathrm{univ}}(0)$ has zero
measure with respect to $|\Psi _{\mathrm{univ}}(q_{\mathrm{univ}},0)|^{2}$
then that same set of points may reasonably be regarded as objectively small
and hence physically negligible. But this objectivity vanishes when we take
the limit $n\rightarrow \infty $ (the limit where the typicality argument is
applied). For example, as we noted above, for $n\rightarrow \infty $ the
nonequilibrium set%
\begin{equation}
S_{\mathrm{noneq}}=\left\{ q_{\mathrm{univ}}(0)\mid \rho (q,0)\propto |\psi
(q,0)|^{4}\right\} 
\end{equation}%
(the set of initial points $q_{\mathrm{univ}}(0)$ yielding a nonequilibrium
sub-system density $\rho (q,0)\propto |\psi (q,0)|^{4}$) has zero measure
with respect to $|\Psi _{\mathrm{univ}}(q_{\mathrm{univ}},0)|^{2}$ (that is, 
$\mu _{\mathrm{eq}}[S_{\mathrm{noneq}}]=0$ where $d\mu _{\mathrm{eq}}=|\Psi
_{\mathrm{univ}}(q_{\mathrm{univ}},0)|^{2}dq_{\mathrm{univ}}$). On the other
hand, the same set $S_{\mathrm{noneq}}$ has unit measure with respect to $%
|\Psi _{\mathrm{univ}}(q_{\mathrm{univ}},0)|^{4}$ (suitably normalised)
(that is, $\mu _{\mathrm{noneq}}[S_{\mathrm{noneq}}]=1$ where $d\mu _{%
\mathrm{noneq}}\propto |\Psi _{\mathrm{univ}}(q_{\mathrm{univ}},0)|^{4}dq_{%
\mathrm{univ}}$). Thus, in the limit $n\rightarrow \infty $, there is 
\textit{no} objective sense in which the nonequilibrium set $S_{\mathrm{noneq%
}}$ is `small'.

Technically, the qualitative difference between cases with finite $n$ and
cases where the literal limit $n\rightarrow\infty$ is taken may be
highlighted in terms of the notion of `absolute continuity'. For finite $n$,
the alternative measure $|\Psi_{\mathrm{univ}}(q_{\mathrm{univ}},0)|^{4}$ is
absolutely continuous with respect to the equilibrium measure $|\Psi _{%
\mathrm{univ}}(q_{\mathrm{univ}},0)|^{2}$. This simply means, by definition,
that the alternative measure of a set $S$ is equal to zero whenever the
equilibrium measure of $S$ is equal to zero ($\mu_{\mathrm{eq}}[S]=0$
implies $\mu_{\mathrm{noneq}}[S]=0$). But this will \textit{not} hold for $%
n\rightarrow\infty$, where we can have both $\mu_{\mathrm{eq}}[S_{\mathrm{%
noneq}}]=0$ and $\mu_{\mathrm{noneq}}[S_{\mathrm{noneq}}]=1$.

In this context one should beware of statements appealing to absolute
continuity. For example, D\"{u}rr and Teufel (2009, p. 222) write:

\begin{quotation}
... any measure ... which is absolutely continuous with respect to the
equivariant measure [$\left\vert \Psi_{\mathrm{univ}}\right\vert ^{2}$] ...
defines the same sense of typicality.
\end{quotation}

\noindent Taken by itself this statement is correct and simply amounts to a
statement of the definition of absolute continuity. But to avoid
misunderstandings in this context, it is important to add (as D\"{u}rr and
Teufel omit to) that absolute continuity fails in the relevant limit $%
n\rightarrow\infty$. In that limit, running the argument with an alternative
typicality measure will \textit{not} yield the Born rule for sub-systems,
because the alternative measure will not be absolutely continuous with
respect to the Born-rule measure.

\begin{center}
\textit{Intrinsic likelihood and unlikelihood}
\end{center}

We have seen that at $t=0$ nonequilibrium is `untypical' (has zero measure)
with respect to the equilibrium measure, and that equally nonequilibrium is
`typical' (has unit measure) with respect to a corresponding nonequilibrium
measure. It might then be said that initial nonequilibrium is unlikely with
respect to the equilibrium measure, and that it is \textit{likely} with
respect to a nonequilibrium measure.

With these clarifications in mind, there is clearly no sense in which
quantum nonequilibrium is instrinsically unlikely -- contrary to claims made
by the Bohmian mechanics school, according to which the theory will always
yield the Born rule. For example, in their founding paper D\"{u}rr \textit{%
et al}. claim that

\begin{quotation}
... in a universe governed by Bohmian mechanics it is in principle
impossible to know more about the configuration of any subsystem than what
is expressed by [the Born rule]. (D\"{u}rr, Goldstein and Zangh\`{\i} 1992)
\end{quotation}

\noindent Similarly, Tumulka writes that

\begin{quotation}
... in a universe governed by Bohmian mechanics, observers will see outcomes
with exactly the probabilities specified by the usual rules of quantum
mechanics ... . (Tumulka 2018)
\end{quotation}

\noindent But as we have noted this is true only in quantum equilibrium: for
general nonequilibrium ensembles the Born rule is violated (Valentini
1991a,b) and it is perfectly possible to know more about a sub-system than
is allowed by the Born rule or by the associated uncertainty principle
(Valentini 2002a).

The claims made by the Bohmian mechanics school amount to an artificial and
unjustified restriction of the theory to equilibrium only. This is most
explicit in the presentation by Tumulka, who writes that it is

\begin{quotation}
... a fundamental law of Bohmian mechanics, to demand that [$q$] be \textit{%
typical} with respect to $\left\vert \Psi\right\vert ^{2}$. (Tumulka 2018)
\end{quotation}

\noindent But as we have emphasised, only the dynamical equations have the
status of fundamental laws. In a theory of dynamics it is physically
nonsensical to regard a restriction on the initial state as a `fundamental
law'.

D\"{u}rr and Teufel (2009, p. 224) go so far as to compare searching for
nonequilibrium to waiting for a stone to jump up spontaneously in the air:

\begin{quotation}
... one could sit in front of a stone and wait for the stone to jump into
the air, because in a very atypical world, that could happen, now, tomorrow,
maybe the day after tomorrow. (D\"{u}rr and Teufel 2009, p. 224)
\end{quotation}

\noindent But there is no scientific basis for the claim that quantum
nonequilibrium is intrinsically unlikely. This claim stems, as we have seen,
from a circular argument in which the Born-rule measure is taken to \textit{%
define} `typicality' for the initial conditions of the universe. In our
view, in contrast, initial conditions are ultimately empirical.

\begin{center}
\textit{Typicality and probability}
\end{center}

If we replace the word `typicality' by its synonym `probability', it becomes
apparent that the argument given by D\"{u}rr \textit{et al}. simply assumes
a universal Born-rule probability density $|\Psi_{\mathrm{univ}}|^{2}$,
which implies the Born-rule probability density $|\psi|^{2}$ for sub-systems.%
\footnote{%
For measurements at a single time this result is a restatement of the
`nesting' property discussed in Section 2 (Valentini 1991a).}

Goldstein (2001) has, however, defended the view that typicality and
probability are \textit{not} synonymous. Goldstein argues that for a set $S$
the precise value of the typicality measure $\mu(S)$ (say 1/2 or 3/4) is
immaterial: the only thing that matters is whether $\mu(S)$ is very small or
not. If $\mu(S)$ is very small, this is regarded as a sufficient explanation
for why events in $S$ do not occur. According to Goldstein, this concept of
typicality plays an important role in scientific explanation. In our view
such attempts to elevate the notion of typicality from a synonym for
probability to a fundamentally new kind of explanatory principle raise more
questions than they answer. It is claimed that what counts is only whether $%
\mu(S)$ is very small or not. How small is small enough? Questions remain as
to precisely how typicality differs from probability. One may also ask why
the notion of probability alone does not suffice. The nature of probability
is already controversial: it seems misguided to introduce a conceptual
variant which proves to be even more controversial.

\begin{center}
\textit{Other responses to circularity}
\end{center}

D\"{u}rr and Teufel (2009, pp. 220--222)\ make some attempt to respond to
our charge of circularity.

The first response appeals to the equivariance of the Born-rule measure over
time. This measure has the special property that its form as a function of $%
\Psi$ is preserved by the dynamics (that is, it is `equivariant'). D\"{u}rr
and Teufel argue that this property singles out the Born-rule measure as the
preferred measure of typicality:

\begin{quotation}
Would another measure ... say the one with density $\left\vert
\Psi\right\vert ^{4}$ ... not yield typicality for the empirical
distribution $\left\vert \varphi\right\vert ^{4}$ by the same argument? In
fact, it would, but only at \textit{exactly} that moment of time where the
measure has the $\left\vert \Psi\right\vert ^{4}$ density. Since the measure
is not equivariant, its density will soon change to something completely
different. ... The equivariant measure of typicality on the other hand is
special ... . Typicality defined by this measure does not depend on time. (D%
\"{u}rr and Teufel 2009, pp. 220--221)
\end{quotation}

\noindent This defence is physically unconvincing. As we have seen, in their
argument for the typicality of the Born rule for sub-systems the universal
Born-rule measure is applied to the initial universal configuration $q_{%
\mathrm{univ}}(0)$ at only one time (the initial time $t=0$). From this we
may immediately derive the Born rule for sub-systems at that same initial
time $t=0$. The time evolution of the measure at $t>0$ is irrelevant to the
statistics of sub-systems at $t=0$. Furthermore, as D\"{u}rr and Teufel
themselves admit, if an alternative universal measure were applied at $t=0$
then by the same derivation we would obtain an alternative (non-Born-rule)
distribution for sub-systems at $t=0$ -- regardless of how the measure
evolves at later times. D\"{u}rr and Teufel wish to argue in favour of
Born-rule typicality for initial configurations at $t=0$, but their argument
appeals to properties of the time evolution of the measure at $t>0$.
Physically speaking, it is hard to understand how initial conditions at $t=0$
can be dictated by, or influenced by, a convenient mathematical property of
the evolution at $t>0$.

In a similar vein, while commenting on the quantum relaxation approach
advocated by this author, and in particular on the idea of primordial
quantum nonequilibrium, Tumulka claims outright that an initial
nonequilibrium measure would be intrinsically unnatural and puzzling because
it would not be equivariant:

\begin{quotation}
... the $\left\vert \Psi\right\vert ^{2}$ distribution is special since it
is equivariant ... if we found empirically (which we have not) that it was
necessary to assume that [$q(0)$] was $\left\vert \Psi\right\vert ^{4}$%
-distributed, then it would be a big puzzle needing explanation why it was $%
\left\vert \Psi\right\vert ^{4}$ of all distributions instead of the
natural, equivariant $\left\vert \Psi\right\vert ^{2}$. (Tumulka 2018)
\end{quotation}

\noindent The argument seems to be that initial equilibrium is natural and
needs no explanation because it is equivariant. Whereas, because initial
nonequilibrium is not equivariant, it is unnatural and so -- if it were
observed -- would present a major puzzle. Again, physically it is difficult
to see how initial conditions at $t=0$ can be dictated by a convenient
mathematical property of the evolution at later times.

A second response to the charge of circularity appeals to ease of proof:

\begin{quotation}
... the equivariant measure is a highly valuable technical tool, because
this is the measure which allows us to prove the theorem! At time $t$, let
us say today when we do the experiment, any other measure would look so odd
(it would depend on $\Psi_{t}$ in such an intricate way) that we would have
no chance of proving anything! And if it did not look odd today, then it
would look terribly odd tomorrow! The equivariant measure always looks the
same and what we prove today about the empirical distribution will hold
forever. (D\"{u}rr and Teufel 2009, p. 222)
\end{quotation}

\noindent But we are concerned with objective physical facts about the
initial configuration of our universe. In a scientific theory, the initial
state of the universe is not determined by mathematical convenience or ease
of proof of a certain theorem, but by empirical observation and measurement.

\begin{center}
\textit{Typicality in classical and quantum statistical mechanics}
\end{center}

In defence of the typicality approach it might be asserted that, in
statistical mechanics generally, certain undesirable initial conditions are
often ruled out on the grounds that they are exceptional (or untypical) with
respect to a particular measure. Some authors do indeed justify the required
restrictions on initial conditions along these lines. But it should be
admitted that all initial conditions are allowed in principle. The actual
realised initial conditions are ultimately an empirical matter to be
constrained by experiment. Furthermore, the mere use of a different word
does not entail the use of a genuinely different concept: in our view ruling
out undesirable initial conditions as `untypical' simply amounts to ruling
them out as `improbable'.

The use of typicality in classical statistical mechanics has also been
defended by Goldstein (2001), who argues that to explain thermal relaxation
it suffices to note that phase-space points corresponding to thermal
equilibrium occupy an overwhelmingly larger volume than phase-space points
corresponding to thermal nonequilibrium. According to this argument, if a
system begins at an initial point corresponding to nonequilibrium, then it
is overwhelmingly likely to evolve (and quickly) to a final point
corresponding to equilibrium, merely by virtue of the much larger volume
occupied by the latter points. It may then be said that thermal relaxation
is `typical' with respect to the phase-space volume measure. However, as
noted by Uffink (2007, pp. 979--980), a specific system trajectory $%
(q(t),p(t))$ traces out a set of points $S_{0}$ of measure zero (with
respect to phase-space volume), whereas the set of points $S_{1}$ never
visited by the trajectory is of measure one. By definition the system
remains in the zero-measure set $S_{0}$ for all time, and does \textit{not}
move into the set $S_{1}$ even though the latter has an overwhelmingly
larger phase-space measure. It is then untenable to claim that a system is
more likely to move into a set of points simply because that set has a
larger phase-space volume. Uffink concludes -- in our view correctly -- that
a \textit{bona fide} explanation for relaxation must appeal to properties of
the dynamics and not merely to a measure-theoretic counting of states.

In our view measure-theoretic arguments are misleading and by themselves
give no indication of likelihood. We emphasise once again that initial
conditions are ultimately a matter for experiment. As scientists we need to
understand which initial conditions are consistent with present
observations. To this end we must consider the dynamics, and also take into
account our knowledge of past history (whether locally in the laboratory or
at cosmological scales).

\subsection{Probability and ensembles in cosmology}

According to the Bohmian mechanics school it is meaningless to consider
probabilities or ensembles for the whole universe:

\begin{quotation}
What physical significance can be assigned to a probability distribution on
the initial configurations for the entire universe? What can be the
relevance to physics of such an ensemble of universes? (D\"{u}rr, Goldstein
and Zangh\`{\i} 1992)

... since we only have access to one universe ... an ensemble of universes
is meaningless for physics. (D\"{u}rr and Teufel 2009, p. 224)
\end{quotation}

\noindent This claim conflicts with common practice in both theoretical and
observational cosmology. In recent decades hundreds of millions of dollars
have been spent on satellite observations of the CMB (as well as on galaxy
surveys) with the express aim of putting empirical constraints on the
probability distribution for primordial cosmological perturbations. These
observations do have a well-defined meaning, in spite of the above claim.

First of all, to suggest that there is only one universe is correct only as
a trivial tautology. In practice cosmologists employ the word `universe' to
denote the totality of what we are currently able to observe. Given our
current knowledge it is perfectly plausible that what we see is only one
element of a very large and perhaps even infinite ensemble. In current
observational cosmology, the data are well described by a `standard'
cosmological model according to which what we see is only a tiny patch
within an infinite flat (expanding)\ space -- a conclusion that is arguably
the most conservative option at the present time. Further afield,
contemporary theoretical cosmology includes `eternal' inflationary models
with an infinity of pocket universes, while string theories are widely
believed to imply the existence of a `multiverse'. Thus in theory there is
no difficulty in imagining that the universe we see is merely one of a large
ensemble, and this may well be the case as a matter of physical fact.

Secondly, the meaning of a probability distribution `for the universe' is by
no means as problematic as the Bohmian mechanics school portrays it. As we
outlined in Section 4, practising cosmologists routinely test the
predictions of such distributions via measurements of the CMB. As we
explained, such measurements probe the primordial power spectrum (\ref{PPS})
for a theoretical ensemble. By this means, whole classes of cosmological
models have been ruled out by observation because they predict an incorrect
spectrum.

This is not to say that one cannot or should not question the meaning of a
theoretical ensemble. But we would argue that such foundational questions
have no special connection with pilot-wave theory or cosmology; they arise
in any practical application of probability theory or statistical inference,
whether one is considering genetic populations on earth or the distribution
of galaxies in space. Furthermore, the relevant mathematical properties of
the assumed `probability distribution over a theoretical ensemble' do not
depend on any particular interpretation of probability theory. One could,
for example, regard the distribution as expressing a subjective degree of
belief, or as representing a really existing ensemble; it would make no
difference to how the distribution is employed in mathematical practice.
Indeed, if one prefers one may avoid the notion of a theoretical ensemble of
`universes' and instead consider a real ensemble of approximately
independent sub-regions of a single universe. Given that the `universe' we
see may in any case be just such a sub-region, which approach one takes is
immaterial.

In this context one should also beware of the claim (sometimes made by the
Bohmian mechanics school\footnote{%
See, for example, Goldstein and Zangh\`{\i} (2013).}) that pilot-wave
dynamics is in some special sense fundamentally a dynamics of the whole
universe. If this were true, we would need a complete theory of cosmology to
work with and apply pilot-wave theory. In a trivial sense, of course, there
will always be small interactions between even very distant systems, and in
all known theories of physics it could be said that fundamentally the theory
is a theory of the whole universe. But this is no more true in pilot-wave
theory than it is in classical mechanics, classical field theory, or general
relativity. In principle one needs to consider the whole universe in all
these theories; but in practice, the universe divides into approximately
independent pieces, at least in the real situations occurring in our actual
world. The situation in pilot-wave theory shows no essential difference from
that of other physical theories.

In our view, the question of the meaning of probability and of ensembles in
cosmology has its place as a valid and interesting philosophical question,
but the emphasis placed on this question by the Bohmian mechanics school has
proved to be misleading and (we would argue) a distraction. The question of
the existence or otherwise of primordial quantum nonequilbrium is empirical.
It will be answered by detailed work in theoretical and observational
cosmology, not by foundational debates about the meaning of probability and
related topics.

On a related note we comment on a recent paper by Norsen (2018), which
attempts to combine quantum relaxation ideas with typicality arguments. In
our view considerations of typicality add nothing substantial to a quantum
relaxation scenario, and merely introduce a new word to denote the
probability measure for a universal theoretical ensemble.\footnote{%
Norsen (p. 15) follows the Bohmian mechanics school in making the misleading
claim that `... there is simply no such thing as the probability
distribution $P$ for particle configurations of the universe as a whole,
because there is just one universe'.} Furthermore if, as Norsen (p. 24)
advocates, we also consider `reasonably smooth' non-Born-rule typicality
measures on the initial universal configuration, then while (as Norsen
notes) our relaxation results suggest that the Born rule will still be
obtained on a coarse-grained level at later times, the fact remains that
such initial measures will imply nonequilibrium for sub-systems in the early
universe (with all the novel physical implications we have described). This
then contradicts Norsen's claim (p. 22) that early nonequilibrium is
intrinsically unlikely (a claim made, confusingly, by appealing to the
initial Born-rule measure). To avoid such needless controversy, we should
bear in mind that there simply is no intrinsic typicality (or probability)
measure for initial conditions, and that in the end the existence or
non-existence of early nonequilibrium can only be established by observation.

\section{Contingency and the nature of the universal wave function}

The typicality approach has led to misunderstandings not only of the Born
rule in pilot-wave theory but also of the nature of the wave function (or
pilot wave).

As we have emphasised, in a theory of dynamics the initial conditions are
contingent and only the laws of motion are law-like. And yet, in the
typicality approach the initial configuration $q_{\mathrm{univ}}(0)$ of the
universe is restricted by the requirement that it be typical with respect to
the measure $|\Psi_{\mathrm{univ}}(q_{\mathrm{univ}},0)|^{2}$. The latter
measure is treated as if it had a law-like status (as explicitly claimed by
Tumulka (2018)). In our view, in contrast, the initial probability
distribution for the universe is a contingency which can be constrained only
by cosmological observation.

This confusion between contingent and law-like entities has been taken to an
extreme in claims made by the Bohmian mechanics school regarding the nature
of the universal wave function $\Psi_{\mathrm{univ}}$, specifically: (1)
that $\Psi_{\mathrm{univ}}$ cannot be regarded as contingent, and (2) that $%
\Psi_{\mathrm{univ}}$ is not a physical object but a law-like entity
(`nomological' rather than `ontological').

The argument that $\Psi_{\mathrm{univ}}$ cannot be regarded as contingent is
essentially this: the wave function for the whole universe

\begin{quotation}
is not controllable: it is what it is. (Goldstein 2010)
\end{quotation}

\noindent There are several problems with this argument. Firstly, the same
could be said of the universal configuration $q_{\mathrm{univ}}$, resulting
in the remarkable conclusion that no properties of our universe can be
regarded as contingent (not even the position of the moon). Secondly, and in
a similar vein, one could just as well say that our universe has only one
spacetime geometry, and indeed only one intergalactic magnetic field. Each
of these objects `is not controllable' and `is what it is'. And yet,
according to standard thinking in physics and cosmology, the detailed form
of either object is not completely determined by physical laws: each has a
strong element of contingency. Thirdly and finally, as noted in Section 5.2,
in this context we ought to beware of statements that there is `only one
universe': in principle such statements are trivially and tautologically
true, but in practice the universe studied by cosmologists may well be one
element of a large (and possibly infinite) ensemble, where the object which
we call $\Psi _{\mathrm{univ}}$ can vary contingently across the ensemble.

The argument that $\Psi_{\mathrm{univ}}$ is not a physical object but a
law-like entity is based on three assertions: (a) that $\Psi_{\mathrm{univ}}$
cannot be regarded as contingent (claim (1)\ above, which we have argued to
be unfounded), (b) that $\Psi_{\mathrm{univ}}$ is static, and (c) that $%
\Psi_{\mathrm{univ}}$ is uniquely determined by the laws of quantum gravity.%
\footnote{%
Other authors express concern about regarding a field on configuration space
as a physical object. In our view it is not unreasonable for configuration
space to be the fundamental arena of a realistic physics, with physical
objects propagating on it (cf. footnote 23).} On these grounds it has been
claimed that $\Psi_{\mathrm{univ}}$ is a law-like entity roughly analogous
to a classical Hamiltonian (D\"{u}rr, Goldstein and Zangh\`{\i} 1997;
Goldstein and Zangh\`{\i} 2013). Thus:

\begin{quotation}
... the wave function is a component of physical law rather than of the
reality described by the law. (D\"{u}rr, Goldstein and Zangh\`{\i} 1997, p.
33)
\end{quotation}

\noindent But the arguments (a)--(c) do not bear scrutiny. We have already
seen that argument (a)\ is spurious. Argument (b) is also questionable,
based as it is on the time-independence of the Wheeler-DeWitt equation (the
analogue of the Schr\"{o}dinger equation) in canonical quantum gravity.%
\footnote{%
The Wheeler-DeWitt equation takes the schematic atemporal form $\mathcal{%
\hat{H}}\Psi =0$, where $\mathcal{\hat{H}}$ is an appropriate operator for
the Hamiltonian density (Rovelli 2004).} However, the physical meaning and
consistency of the quantum-gravitational formalism remains in doubt, in
particular because of the notorious `problem of time' (the problem of
explaining the emergence of apparent temporal evolution in some appropriate
limit). Many workers have suggested that a physical time parameter is in
effect hidden within the formalism and that, when correctly written as a
function of physical degrees of freedom, the wave function is in fact time
dependent.\footnote{%
See, for example, Roser and Valentini (2014) and the exhaustive review of
the problem of time by Anderson (2017).} As for argument (c), in canonical
quantum gravity the solutions for $\Psi _{\mathrm{univ}}$ (satisfying the
Wheeler-DeWitt equation as well as the other required constraints) are in
fact highly \textit{non}-unique (Rovelli 2004). In quantum cosmological
models, for example, the solutions for $\Psi _{\mathrm{univ}}$ have the same
kind of contingency that we are used to for quantum states in other areas of
physics (Bojowald 2015).

It is worth emphasising that even if a consistent theory of quantum gravity
did require $\Psi_{\mathrm{univ}}$ to be static, this still would not by any
means establish that $\Psi_{\mathrm{univ}}$ is law-like. The key aspect of $%
\Psi_{\mathrm{univ}}$ that makes it count as a physical object is its
contingency, in other words its under-determination by known physical laws.
This implies that $\Psi_{\mathrm{univ}}$ contains a lot of independent and
contingent structure -- just like the electromagnetic field or the universal
spacetime geometry -- and so should be regarded as part of the physical
state of the world (Valentini 1992, p. 17; Brown and Wallace 2005, p. 532;
Valentini 2010b).

\section{\textbf{Further criticisms of `Bohmian mechanics'}}

Pilot-wave theory is, in general, a nonequilibrium physics that violates the
statistical predictions of quantum theory (Valentini 1991a,b, 1992). It can
only be properly understood from this general perspective. The Bohmian
mechanics school has instead promoted the belief that pilot-wave theory is
instrinsically a theory of equilibrium. We now consider the principal
physical misunderstandings that have arisen from this mistaken belief.

\subsection{`Absolute uncertainty'}

The Bohmian mechanics school has asserted that typicality with respect to
the Born-rule measure is `the origin of absolute uncertainty'. On this view
the uncertainty principle is an `absolute' and `irreducible' limitation on
our knowledge:

\begin{quotation}
In a universe governed by Bohmian mechanics there are sharp, precise, and
irreducible limitations on the possibility of obtaining knowledge ...
absolute uncertainty arises as a necessity, emerging as a remarkably clean
and simple consequence of the existence of trajectories. (D\"{u}rr,
Goldstein and Zangh\`{\i} 1992)

... in a Bohmian universe we have an \textit{absolute uncertainty} ... the
[Born rule] is a sharp expression of the inaccessibility in a Bohmian
universe of micro-reality, of the unattainability of knowledge of the
configuration of a system that transcends the limits set by its wave
function $\psi$. (Goldstein 2010)

In Bohmian mechanics ... there are sharp limitations to knowledge and
control: inhabitants of a Bohmian universe cannot know the position of a
particle more precisely than allowed by the $\left\vert \psi\right\vert ^{2}$
distribution ... . Furthermore, they cannot measure the position at time $t$
without disturbing the particle ... . (Tumulka 2018)
\end{quotation}

\noindent But as we pointed out in Section 5.1, the typicality argument in
effect inserts the Born rule by hand at the initial time. Furthermore, as we
noted in the Introduction, the uncertainty principle is not absolute or
irreducible but merely a peculiarity of the state of quantum equilibrium. In
general, the uncertainty principle would be violated if we had access to
quantum nonequilibrium systems (Valentini 1991b, 2002a).

\subsection{The status of quantum measurement theory}

As also briefly noted in the Introduction, in the presence of quantum
nonequilibrium key quantum constraints are violated: these include
statistical locality, expectation additivity for quantum observables, and
the indistinguishability of non-orthogonal quantum states (Valentini
1991a,b, 1992, 2002a, 2004; Pearle and Valentini 2006). The physics of
nonequilibrium is radically different from the physics of equilibrium, the
latter being merely a highly restricted special case of the former. It
should then come as no surprise that the nonequilibrium theory of
measurement differs radically from its equilibrium counterpart. As
philosophers of physics are well aware, measurement is `theory laden': we
need some body of theory in order to know how to perform measurements
correctly. As Einstein put it, in an often-quoted conversation with
Heisenberg:

\begin{quotation}
It is the theory which decides what we can observe. (Heisenberg 1971, p. 63)
\end{quotation}

In the presence of quantum nonequilibrium systems, pilot-wave theory itself
tells us how to perform correct measurements. It is found, for example, that
if we possessed an ensemble of `apparatus pointers' with an arbitrarily
narrow nonequilibrium distribution (much narrower than the standard quantum
width as defined by the initial wave function of the pointer), then it would
be possible to use the apparatus to perform `subquantum measurements': in
particular, we would be able to measure the position and trajectory of a
particle without disturbing its wave function (to arbitrary accuracy)
(Valentini 2002a, Pearle and Valentini 2006).

From this perspective, the physics of quantum equilibrium is highly
misleading -- and so is the associated equilibrium theory of measurement
(also known as `quantum measurement theory'). In fact, the detailed dynamics
of pilot-wave theory shows that the procedures known as `quantum
measurements' are generally not correct measurements. Instead, those
procedures are merely special kinds of experiments which have been designed
to respect a formal analogy with classical measurements (where the analogy
is implemented by a mathematical correspondence between classical and
quantum Hamiltonians) (Valentini 1992, 1996, 2010b).

To put the so-called quantum theory of `measurement' in a proper
perspective, we must consider the more general physics of quantum
nonequilibrium and its associated theory of subquantum measurement. But
because the Bohmian mechanics school believes that the theory is
fundamentally grounded in equilibrium, they are led to believe that the
equilibrium theory \textit{is} the theory -- and that the associated quantum
theory of measurement has a fundamental status. Thus D\"{u}rr, Goldstein and
Zangh\`{\i} (1996, 2004) argue that the quantum theory of measurement arises
as an account of what they call `reproducible experiments' and reproducible
`measurement-like' experiments. Measurements that lie outside of the domain
of the quantum formalism are not considered. Thus both quantum equilibrium
and its associated theory of measurement are in effect regarded as
fundamental features of pilot-wave theory. In our view this is deeply
mistaken. The physics of equilibrium is a special case of a much wider
physics in which new kinds of measurements are possible. If instead we
artificially restrict ourselves to the equilibrium domain, the result is a
distorted understanding of measurement and an overstatement of the
significance of the conventional quantum formalism.

\subsection{The misleading kinematics of quantum equilibrium}

It is worth noting how the artificial restriction to quantum equilibrium
makes the idea of fundamental Lorentz invariance (at the level of the
underlying equations of motion) seem much more plausible than it really is.
As we have remarked, in general nonequilibrium gives rise to instantaneous
signaling between remote entangled systems (Valentini 1991b).\footnote{%
Similar conclusions hold in all nonlocal and deterministic hidden-variables
theories (Valentini 2002b).} The reality in principle of superluminal
communication between widely-separated experimenters strongly suggests the
existence of an absolute simultaneity associated with a preferred slicing of
spacetime (Valentini 2008b). And indeed most versions of pilot-wave dynamics
(and in particular of quantum field theory) are defined with respect to a
preferred frame with a preferred time parameter $t$ -- where effective
Lorentz invariance emerges only at the statistical level of quantum
equilibrium (Bohm, Hiley and Kaloyerou 1987, Valentini 1992, Bohm and Hiley
1993, Holland 1993). If instead the theory is always and everywhere
artificially restricted to equilibrium, locality will always hold at the
statistical level and practical nonlocal signalling will be impossible. It
may then seem plausible to search for a version of pilot-wave theory in
which the dynamics is fundamentally Lorentz invariant, since one will never
be faced directly with the awkward question of what happens when practical
superluminal signals are viewed from a Lorentz-boosted frame and appear to
travel backwards in time (potentially generating causal paradoxes). Even so,
despite several attempts, a fundamentally Lorentz-invariant pilot-wave
theory remains elusive and problematic (D\"{u}rr \textit{et al}. 1999;
Tumulka 2007; D\"{u}rr \textit{et al}. 2014).

The attachment to fundamental Lorentz invariance has in turn encouraged a
misunderstanding of the role of Galilean invariance, which the Bohmian
mechanics school mistakenly regards as a fundamental symmetry of the
low-energy theory (D\"{u}rr, Goldstein and Zangh\`{\i} 1992; Allori \textit{%
et al}. 2008; D\"{u}rr and Teufel 2009; Goldstein 2017; Tumulka 2018).
Pilot-wave theory is a first-order or `Aristotelian' dynamics with a law of
motion (\ref{deB}) for velocity (as first envisaged by de Broglie in 1923),
in contrast with Newtonian theory which is a second-order dynamics with a
law of motion for acceleration. Because of this fundamental difference, the
natural kinematics of pilot-wave theory is also Aristotelian with a
preferred state of rest (Valentini 1997). Galilean invariance may be shown
to be a fictitious symmetry of the low-energy pilot-wave theory of particles
-- just as invariance under uniform acceleration is well known to be a
fictitious symmetry of Newtonian mechanics. If instead one tries to insist
on Galilean invariance being a physical symmetry of the low-energy theory,
the result is a conceptually incoherent combination of an Aristotelian
dynamics with a Galilean kinematics.

In response it might be claimed that Galilean invariance plays an important
role in selecting the form of the low-energy guidance equation (D\"{u}rr,
Goldstein and Zangh\`{\i} 1992; D\"{u}rr and Teufel 2009; Goldstein 2017).
But in fact the de Broglie velocity $v=j/\left\vert \psi \right\vert ^{2}$
is generally determined by the quantum current $j$, which may be derived as
a Noether current associated with a global phase symmetry $\psi \rightarrow
\psi e^{i\theta }$ on configuration space (Struyve and Valentini 2009). The
derivation takes place in one frame of reference, with no need to consider
boosts. The relevant symmetry is in configuration space, not in space or
spacetime.\footnote{%
This reinforces our view that configuration space is the fundamental
physical arena of pilot-wave dynamics (cf. footnote 19).}

\subsection{Particle creation and indeterminism}

It is also worth noting how the artificial restriction to quantum
equilibrium makes a fundamentally stochastic model of particle creation --
developed by the Bohmian mechanics school -- appear more plausible than it
really is. For if one denies the general contingency of the Born rule for
initial conditions, it may seem no great loss to introduce a fixed and
non-contingent probability into the dynamics as well.

The stochastic model promoted by the Bohmian mechanics school was
constructed as follows. Bell (1986; 1987, chapter 19) had already proposed a
discrete model of fermion numbers evolving stochastically on a lattice and
had suggested that taking the continuum limit might yield a deterministic
theory. The Bohmian mechanics school studied the continuum limit of Bell's
model and arrived at a theory of particle trajectories with stochastic jumps
at events where the particle numbers change (D\"{u}rr \textit{et al}. 2004,
2005). They named their approach `Bell-type quantum field theory', and have
attempted to apply it to bosons as well as to fermions. The fundamental
probability rule for the jumps is chosen so as to preserve the Born rule.

Any interacting quantum field theory will contain a plethora of events where
the particle numbers change (photon emission, electron-positron pair
creation, and so on), and the Bohmian mechanics school has suggested that
determinism must be abandoned to describe them:

\begin{quotation}
In Bell-type [quantum field theories], God does play dice. There are no
hidden variables which would fully predetermine the time and destination of
a jump. (D\"{u}rr \textit{et al}. 2004, p. 3)

The quantum equilibrium distribution, playing a central role in Bohmian
mechanics, then more or less dictates that creation of a particle occurs in
a stochastic manner ... . (D\"{u}rr \textit{et al}. 2005, p. 2)
\end{quotation}

\noindent It would, however, be remarkable indeed if indeterminism were
required to describe particle creation -- when determinism suffices to
describe all other quantum-mechanical processes. But in fact, indeterminism
is not required. The Bohmian mechanics school obtained a\ stochastic
continuum limit of Bell's model because they adopted an erroneous definition
of fermion number $F$. In quantum field theory, $F$ is conventionally
defined as the number of particles minus the number of anti-particles.%
\footnote{%
In particle physics, for historical reasons $F$ is defined as the sum $F=L+B$
of lepton and baryon numbers -- where $L$ is the number of leptons minus the
number of antileptons and similarly for $B$.} As a particle physicist, this
is what Bell would have meant by fermion number. Unfortunately, D\"{u}rr 
\textit{et al}. mistakenly took Bell's `fermion number' to mean the number
of particles \textit{plus} the number of anti-particles.

The correct continuum limit of Bell's model was taken by Colin (2003), who
employed the standard definition of $F$. As a result Colin obtained a `Dirac
sea' theory of fermions -- anticipated by Bohm and Hiley (1993, p. 276) --
in which particle trajectories are determined by a pilot wave that obeys the
many-body Dirac equation.\footnote{%
The Dirac-sea model requires regularisation (for example a cutoff) (Colin
2003, Colin and Struyve 2007). The same is true of the models developed by D%
\"{u}rr \textit{et al}. in the presence of interactions.} The resulting
model is fully deterministic, as Bell suggested it would be. There are no
fixed or fundamental stochastic elements, and the usual contingency of
probabilities applies to all processes.

For completeness we note that, for bosons, it is straightforward to develop
a deterministic pilot-wave field theory, in which the time evolution of a
(for example scalar) field $\phi$ is determined by the Schr\"{o}dinger wave
functional $\Psi\lbrack\phi,t]$ (Holland 1993). In such a theory, again, the
Born rule $P=\left\vert \Psi\right\vert ^{2}$ is contingent and may be
understood as arising from a process of dynamical relaxation (Valentini
2007). In contrast, the Bohmian mechanics school encounters difficulties
defining particle trajectories and a Born-rule position-space density for
single bosons, as briefly noted by D\"{u}rr \textit{et al}. (2005, p. 13).
Such problems recall the long history (in standard quantum theory) of
controversial attempts to define a position-space `wave function' for single
photons and other bosons, attempts which invariably lead to negative
probabilities and superluminal wave packet propagation. Without a solution
to this -- probably insoluble -- problem, so-called `Bell-type quantum field
theory' remains undefined for bosons.\footnote{%
D\"{u}rr \textit{et al}. (2005, p. 13) cite two papers `in preparation'
(their refs. [18] and [28]) purporting to address this problem. To the
author's knowledge, and unsurprisingly, neither paper was completed.}

\subsection{The problem of falsifiability}

Finally, the artificial restriction to quantum equilibrium has compromised
the status of pilot-wave theory as a falsifiable scientific theory. For it
is then impossible to measure the trajectory of a system without disturbing
its wave function; hence it is impossible to test the de Broglie equation of
motion, which associates a specific set of trajectories with each given wave
function. There are alternative pilot-wave theories, with alternative
velocity fields, which nevertheless preserve the Born distribution and which
therefore imply the same empirical predictions for the equilibrium state
(Deotto and Ghirardi 1998). The physics of equilibrium is insensitive to the
details of the trajectories. Thus, if we have access to equilibrium only,
the alternative theories can never be tested against de Broglie's original
theory. Indeed, in equilibrium, pilot-wave theories are forever
experimentally indistinguishable from conventional quantum theory. As D\"{u}%
rr \textit{et al}. put it:

\begin{quotation}
For every conceivable experiment, whenever quantum mechanics makes an
unambiguous prediction, Bohmian mechanics makes exactly the same prediction.
Thus, the two cannot be tested against each other. (D\"{u}rr, Goldstein,
Tumulka and Zangh\`{\i} 2009)
\end{quotation}

It is of course logically and mathematically possible for the world to be
governed by pilot-wave theory and to be always and everywhere in quantum
equilibrium. But from a scientific point of view such a theory is
unfalsifiable and therefore unacceptable. This shortcoming is, however, not
a feature of pilot-wave theory itself -- which abounds in new and
potentially-observable physics -- but rather stems from a misunderstanding
of the status of the Born rule in this theory.

\section{Conclusion}

The foundations of statistical mechanics are notoriously controversial, and
overlap with difficult questions concerning the nature of probability, the
justification for standard methods of statistical inference, and even with
philosophical questions concerning the foundations of the scientific method.
However, important as these questions are, in our view they are not
especially relevant to either pilot-wave theory or cosmology but instead
arise generally across the sciences. We claim that attempts to forge a
special link between such questions and pilot-wave theory are at best a
distraction and at worst deeply misleading.

Something comparable took place during the early development of atomic
theory in the late nineteenth century. At that time theoretical physics was
divided between what we might now call `operationalists' (who saw the
macroscopic laws of thermodynamics as paradigmatic for physics generally)
and `realists' (who thought those laws required a deeper explanation in
terms of atoms and kinetic theory). Boltzmann in particular was especially
passionate about the philosophical importance of atomism as a basis for
explanation in physics, vis \`{a} vis the competing operationalist views of
Mach and Ostwald.\footnote{%
See, for example, Boltzmann's selected writings in the collection \textit{%
Theoretical Physics and Philosophical Problems} (Boltzmann 1974).} In
retrospect it seems unfortunate that Boltzmann became embroiled in
foundational controversies concerning probability, time reversal, and so on
-- important and interesting questions which, in hindsight, proved to be a
distraction from the main goal of demonstrating the reality of atoms. The
eventual atomistic explanation for Brownian motion by Einstein in 1905 owed
little to such foundational debates and more to technical developments in
kinetic theory -- and the foundational debates persist to this day, more
than a century after the existence of atoms was firmly established.

Similarly, theoretical physics today is again divided between
operationalists (who see quantum mechanics as an operational theory of
macroscopic observations) and realists (who think there must be a reality
behind the formalism). Among the various realist approaches, pilot-wave
theory is the most closely analogous to kinetic theory. Once again, it seems
unfortunate that the subject has become embroiled in foundational
controversies in statistical mechanics and probability theory, when surely
the main goal is to find out whether the trajectories posited by pilot-wave
theory really exist or not. In our view, while those foundational
controversies are important and interesting, in the context of pilot-wave
theory they have proved to be a distraction from the even more important
question of whether pilot-wave theory itself is true or not. Furthermore,
the viewpoint championed by the Bohmian mechanics school (and widely
followed by philosophers of physics) has played a major role in obscuring
the physics of the theory -- which is fundamentally a nonequilibrium physics
that violates quantum mechanics. We emphasise, once again, that the
existence or non-existence of quantum nonequilibrium in our universe (past
and present) is an empirical question that will be settled only by detailed
theoretical and observational work.

\textbf{Acknowledgement}. I am grateful to Valia Allori for the invitation
to contribute to this volume.

\begin{center}
\textbf{BIBLIOGRAPHY}
\end{center}

Abraham, E., Colin, S. and Valentini, A. (2014). Long-time relaxation in
pilot-wave theory. \textit{Journal of Physics A}, 47: 395306.
[arXiv:1310.1899]

Aghanim, N. \textit{et al}. (Planck Collaboration) (2016). \textit{Planck}
2015 results. XI. CMB power spectra, likelihoods, and robustness of
parameters. \textit{Astronomy and Astrophysics}, 594: A11. [arXiv:1507.02704]

Allori, V., Goldstein, S., Tumulka, R. and Zangh\`{\i}, N. (2008). On the
common structure of Bohmian mechanics and the Ghirardi--Rimini--Weber
theory.\ \textit{British Journal for the Philosophy of Science}, 59:
353--389. [arXiv:quant-ph/0603027]

Anderson, E. (2017). \textit{The Problem of Time}. Springer-Verlag.

Bacciagaluppi, G. and Valentini, A. (2009). \textit{Quantum Theory at the
Crossroads: Reconsidering the 1927 Solvay Conference}. Cambridge: Cambridge
University Press. [arXiv:quant-ph/0609184]

Bell, J. S. (1986). Quantum field theory without observers. \textit{Physics
Reports}, 137: 49--54.

Bell, J. S. (1987). \textit{Speakable and Unspeakable in Quantum Mechanics}.
Cambridge: Cambridge University Press.

Bohm, D. (1952a). A suggested interpretation of the quantum theory in terms
of `hidden' variables. I. \textit{Physical Review}, 85: 166--179.

Bohm, D. (1952b). A suggested interpretation of the quantum theory in terms
of `hidden' variables. II. \textit{Physical Review}, 85: 180--193.

Bohm, D. and Hiley, B. J. (1993). \textit{The Undivided Universe: an
Ontological Interpretation of Quantum Theory}. Routledge.

Bohm, D., Hiley, B. J. and Kaloyerou, P. N. (1987). An ontological basis for
the quantum theory. \textit{Physics Reports}, 144: 321--375.

Bojowald, M. (2015). Quantum cosmology: a review. \textit{Reports on
Progress in Physics}, 78: 023901. [arXiv:1501.04899]

Boltzmann, L. (1974). \textit{Theoretical Physics and Philosophical Problems}%
, ed. B. McGuinness. Dordrecht: Reidel.

Colin, S. (2003). A deterministic Bell model. \textit{Physics Letters A},
317: 349--358. [arXiv:quant-ph/0310055]

Colin, S. (2012). Relaxation to quantum equilibrium for Dirac fermions in
the de Broglie-Bohm pilot-wave theory. \textit{Proceedings of the Royal
Society A}, 468: 1116--1135. [arXiv:1108.5496]

Colin, S. and Struyve, W. (2007). A Dirac sea pilot-wave model for quantum
field theory. \textit{Journal of Physics A}, 40: 7309--7341.
[arXiv:quant-ph/0701085]

Colin, S. and Valentini, A. (2013). Mechanism for the suppression of quantum
noise at large scales on expanding space. \textit{Physical Review D}, 88:
103515. [arXiv:1306.1579]

Colin, S. and Valentini, A. (2014). Instability of quantum equilibrium in
Bohm's dynamics. \textit{Proceedings of the Royal Society A}, 470: 20140288.

Colin, S. and Valentini, A. (2015). Primordial quantum nonequilibrium and
large-scale cosmic anomalies. \textit{Physical Review D}, 92: 043520.
[arXiv:1407.8262]

Colin, S. and Valentini, A. (2016). Robust predictions for the large-scale
cosmological power deficit from primordial quantum nonequilibrium. \textit{%
International Journal of Modern Physics D}, 25: 1650068. [arXiv:1510.03508]

de Broglie, L. (1928). La nouvelle dynamique des quanta. In \textit{\'{E}%
lectrons et Photons: Rapports et Discussions du Cinqui\`{e}me Conseil de
Physique}. Paris: Gauthier-Villars, pp. 105--132. [English translation:
Bacciagaluppi, G. and Valentini, A. (2009).]

Deotto, E and Ghirardi, G. C. (1998). Bohmian mechanics revisited. \textit{%
Foundations of Physics}, 28: 1--30. [arXiv:quant-ph/9704021]

D\"{u}rr, D., Goldstein, S., and Zangh\`{\i}, N. (1992). Quantum equilibrium
and the origin of absolute uncertainty. \textit{Journal of Statistical
Physics}, 67: 843--907. [arXiv:quant-ph/0308039]

D\"{u}rr, D., Goldstein, S., and Zangh\`{\i}, N. (1996). Bohmian mechanics
as the foundation of quantum mechanics. In \textit{Bohmian Mechanics and
Quantum Theory: an Appraisal}, eds. J. T. Cushing, A. Fine and S. Goldstein.
Dordrecht: Kluwer, pp. 21--44. [arXiv:quant-ph/9511016]

D\"{u}rr, D., Goldstein, S., and Zangh\`{\i}, N. (1997). Bohmian mechanics
and the meaning of the wave function. In \textit{Experimental Metaphysics:
Quantum Mechanical Studies for Abner Shimony}, eds. R. S. Cohen, M. Horne,
and J. Stachel. Dordrecht: Kluwer, pp. 25--38. [arXiv:quant-ph/9512031]

D\"{u}rr, D., Goldstein, S., M\"{u}nch-Berndl, K., and Zanghi, N. (1999).
Hypersurface Bohm-Dirac models. \textit{Physical Review A}, 60: 2729--2736.

D\"{u}rr, D., Goldstein, S., and Zangh\`{\i}, N. (2004). Quantum equilibrium
and the role of operators as observables in quantum theory. \textit{Journal
of Statistical Physics}, 116: 959--1055. [arXiv:quant-ph/0308038]

D\"{u}rr, D., Goldstein, S., Tumulka, R., and Zangh\`{\i}, N. (2004).
Bohmian mechanics and quantum field theory. \textit{Physical Review Letters}%
, 93: 090402. [arXiv:quant-ph/0303156]

D\"{u}rr, D., Goldstein, S., Tumulka, R., and Zangh\`{\i}, N. (2005).
Bell-type quantum field theories. \textit{Journal of Physics A}, 38:
R1--R43. [arXiv:quant-ph/0407116]

D\"{u}rr, D., Goldstein, S., Tumulka, R., and Zangh\`{\i}, N. (2009).
Bohmian Mechanics. In \textit{Compendium of Quantum Physics: Concepts,
Experiments, History and Philosophy}, eds. D. Greenberger, K. Hentschel and
F. Weinert. Springer-Verlag, pp. 47--55. [arXiv:0903.2601]

D\"{u}rr, D., Goldstein, S., Norsen, T., Struyve, W. and Zangh\`{\i}, N.
(2014). Can Bohmian mechanics be made relativistic? \textit{Proceedings of
the Royal Society A}, 470: 20130699. [arXiv:1307.1714]

D\"{u}rr, D. and Teufel, S. (2009). \textit{Bohmian Mechanics: The Physics
and Mathematics of Quantum Theory}. Springer-Verlag.

Gibbs, J. W. (1902). \textit{Elementary Principles in Statistical Mechanics}%
. New York: Charles Scribner's Sons.

Goldstein, S. (2001). Boltzmann's approach to statistical mechanics. In 
\textit{Chance in Physics: Foundations and Perspectives}, eds. J. Bricmont 
\textit{et al}.. Springer-Verlag. [arXiv:cond-mat/0105242]

Goldstein, S. (2010). Bohmian mechanics and quantum information. \textit{%
Foundations of Physics}, 40: 335--355. [arXiv:0907.2427]

Goldstein, S. (2017). Bohmian mechanics. In \textit{The Stanford
Encyclopedia of Philosophy}, ed. E. N. Zalta (Summer 2017 Edition).

[https://plato.stanford.edu/archives/sum2017/entries/qm-bohm]

Goldstein, S., Struyve, W., and Tumulka, R. (2015). The Bohmian approach to
the problems of cosmological quantum fluctuations. ArXiv:1508.01017.

Goldstein, S. and Zangh\`{\i}, N. (2013). Reality and the role of the
wavefunction in quantum theory. In \textit{The Wave Function: Essays in the
Metaphysics of Quantum Mechanics}, eds. D. Albert and A. Ney. Oxford: Oxford
University Press, pp. 91--109. [arXiv:1101.4575]

Hajian, A. and Souradeep, T. (2005). The cosmic microwave background bipolar
power spectrum: basic formalism and applications. ArXiv:astro-ph/0501001.

Heisenberg, W. (1971). \textit{Physics and Beyond}. New York: Harper{\&}Row.

Holland, P. R. (1993). \textit{The Quantum Theory of Motion: an Account of
the de Broglie-Bohm Causal Interpretation of Quantum Mechanics}. Cambridge:
Cambridge University Press.

Kandhadai, A. and Valentini, A. (2019). In preparation.

Liddle, A. R. and Lyth, D. H. (2000). \textit{Cosmological Inflation and
Large-Scale Structure}. Cambridge: Cambridge University Press.

Lyth, D. H. and Riotto, A. (1999). Particle physics models of inflation and
the cosmological density perturbation. \textit{Physics Reports}, 314: 1--146.

Norsen, T. (2018). On the explanation of Born-rule statistics in the de
Broglie-Bohm pilot-wave theory. \textit{Entropy}, 20: 422.

Pearle, P. and Valentini, A. (2006). Quantum mechanics: generalizations. In 
\textit{Encyclopaedia of Mathematical Physics}, eds. J.-P. Fran\c{c}oise 
\textit{et al}.. North-Holland: Elsevier. [arXiv:quant-ph/0506115]

Roser, P. and Valentini, A. (2014). Classical and quantum cosmology with
York time. \textit{Classical and Quantum Gravity}, 31: 245001.
[arXiv:1406.2036]

Rovelli, C. (2004). \textit{Quantum Gravity}. Cambridge: Cambridge
University Press.

Struyve, W. and Valentini, A. (2009). De Broglie-Bohm guidance equations for
arbitrary Hamiltonians. \textit{Journal of Physics A}, 42: 035301.
[arXiv:0808.0290]

Towler, M. D., Russell, N. J., and Valentini, A. (2012). Time scales for
dynamical relaxation to the Born rule. \textit{Proceedings of the Royal
Society A}, 468: 990--1013. [arXiv:1103.1589]

Tumulka, R. (2007). The `unromantic pictures' of quantum theory.\textit{\
Journal of Physics A}, 40: 3245--3273. [arXiv:quant-ph/0607124]

Tumulka, R. (2018). Bohmian mechanics. In \textit{The Routledge Companion to
the Philosophy of Physics}, eds. E. Knox and A. Wilson. New York: Routledge.
[arXiv:1704.08017]

Uffink, J. (2007). Compendium of the foundations of classical statistical
physics. In \textit{Philosophy of Physics (Handbook of the Philosophy of
Science)}, eds. J. Butterfield and J. Earman. Amsterdam: North-Holland.

Underwood, N. G. and Valentini, A. (2015). Quantum field theory of relic
nonequilibrium systems. \textit{Physical Review D}, 92: 063531.
[arXiv:1409.6817]

Underwood, N. G. and Valentini, A. (2016). Anomalous spectral lines and
relic quantum nonequilibrium. ArXiv:1609.04576.

Valentini, A. (1991a). Signal-locality, uncertainty, and the subquantum
H-theorem. I. \textit{Physics Letters A}, 156: 5--11.

Valentini, A. (1991b). Signal-locality, uncertainty, and the subquantum
H-theorem, II. \textit{Physics Letters A}, 158: 1--8.

Valentini, A. (1992). On the pilot-wave theory of classical, quantum and
subquantum physics. Ph.D. thesis, International School for Advanced Studies,
Trieste, Italy. [http://hdl.handle.net/20.500.11767/4334]

Valentini, A. (1996). Pilot-wave theory of fields, gravitation and
cosmology. In \textit{Bohmian mechanics and quantum theory: an appraisal},
eds. J. T. Cushing, A. Fine, and S. Goldstein. Dordrecht: Kluwer.

Valentini, A. (1997). On Galilean and Lorentz invariance in pilot-wave
dynamics. \textit{Physics Letters A}, 228: 215. [arXiv:0812.4941]

Valentini, A. (2001). Hidden variables, statistical mechanics and the early
universe. In \textit{Chance in Physics: Foundations and Perspectives}, eds.
J. Bricmont \textit{et al}.. Springer-Verlag. [arXiv:quant-ph/0104067]

Valentini, A. (2002a). Subquantum information and computation. \textit{%
Pramana---Journal of Physics}, 59: 269--277. [arXiv:quant-ph/0203049].

Valentini, A. (2002b). Signal-locality in hidden-variables theories. \textit{%
Physics Letters A}, 297: 273--278. [arXiv:quant-ph/0106098]

Valentini, A. (2004). Universal signature of non-quantum systems. \textit{%
Physics Letters A}, 332: 187--193. [arXiv:quant-ph/0309107]

Valentini, A. (2007). Astrophysical and cosmological tests of quantum
theory. \textit{Journal of Physics A}, 40: 3285--3303. [arXiv:hep-th/0610032]

Valentini, A. (2008a). De Broglie-Bohm prediction of quantum violations for
cosmological super-Hubble modes. ArXiv:0804.4656.

Valentini, A. (2008b). Hidden variables and the large-scale structure of
space-time. In \textit{Einstein, Relativity and Absolute Simultaneity}, eds.
W. L. Craig and Q. Smith. London: Routledge, pp. 125--155.
[arXiv:quant-ph/0504011]

Valentini, A. (2009). Beyond the quantum. \textit{Physics World} 22N11:
32--37. [arXiv:1001.2758]

Valentini, A. (2010a). Inflationary cosmology as a probe of primordial
quantum mechanics. \textit{Physical Review D}, 82: 063513. [arXiv:0805.0163]

Valentini, A. (2010b). De Broglie-Bohm pilot-wave theory: many-worlds in
denial? In \textit{Many Worlds? Everett, Quantum Theory, and Reality}, eds.
S. Saunders \textit{et al}.. Oxford: Oxford University Press, pp. 476--509.
[arXiv:0811.0810]

Valentini, A. (2015). Statistical anisotropy and cosmological quantum
relaxation. ArXiv:1510.02523.

Valentini, A. and Westman, H. (2005). Dynamical origin of quantum
probabilities. \textit{Proceedings of the Royal Society A}, 461: 253.
[arXiv:quant-ph/0403034]

Vitenti, S., Peter, P., and Valentini, A. (2019). Modeling the large-scale
power deficit. ArXiv:1901.08885.

Wang, I.-C. and Ng, K.-W. (2008). Effects of a preinflation
radiation-dominated epoch to CMB anisotropy. \textit{Physical Review D}, 77:
083501.

\end{document}